\documentclass[aps,prd,superscriptaddress,preprintnumbers,twocolumn]{revtex4}

\usepackage{todonotes}
\usepackage{epsf,epsfig,amsmath,amssymb,amsfonts,hyperref,framed,
	microtype,ytableau,bbold,comment,mathrsfs,bm}
\usepackage{xcolor}
\usepackage{slashed}
\usepackage{booktabs}
\usepackage{tikz-cd,tikz}
\usetikzlibrary{arrows,decorations.markings}

\hypersetup{
	pdftitle={plane partitions},
	pdfsubject={High Energy Physics},
	pdfauthor={Thiago Araujo},
	pdfkeywords={gauge; susy; strings; fields; cft},
	pdfsubject={hep-th},
	colorlinks=true,linkcolor=link,citecolor=link,urlcolor=link,linktocpage
}

\definecolor{MyDarkBlue}{rgb}{0.15,0.15,0.45}
\definecolor{shadecolor}{rgb}{0.85,0.85,0.85}
\definecolor{link-blue}{rgb}{0.15,0.15,0.65}
\definecolor{link-red}{rgb}{0.8,0.15,0.1}
\definecolor{link-green}{rgb}{0.15,0.50,0.15}
\definecolor{link}{rgb}{0.45,0.18,0.22}

\newcommand{\be}{\begin{equation}}
\newcommand{\ee}{\end{equation}}	
\newcommand{\bse}{\begin{subequations}}
\newcommand{\ese}{\end{subequations}}
\def\normord{ {\scriptstyle {{\bullet}\atop{\bullet}}} }
\newcommand{\rrangle}{\rangle\hspace{-2.5pt}\rangle}

\def\spp{\tikz[baseline=-.5ex]{
		\fill (-1ex,0) circle (1pt) coordinate (A1);
		\fill (-.4ex,.8ex) circle (1pt) coordinate (A2);
		\fill (.4ex,.8ex) circle (1pt) coordinate (B1);
		\fill (1ex,0) circle (1pt) coordinate (B2);
		\fill (-.5ex,-.5ex) circle (1pt) coordinate (C1);
		\fill (.5ex,-.5ex) circle (1pt) coordinate (C2);
		\draw (A1)--(A2); \draw (B1)--(B2); \draw (C1)--(C2);}
}
\def\blo{\tikz[baseline=-.5ex]{
		\fill (-1ex,0) circle (1pt) coordinate (A1);
		\fill (-.4ex,.8ex) circle (1pt) coordinate (A2);
		\fill (.4ex,.8ex) circle (1pt) coordinate (B1);
		\fill (1ex,0) circle (1pt) coordinate (B2);
		\fill (-.5ex,-.5ex) circle (1pt) coordinate (C1);
		\fill (.5ex,-.5ex) circle (1pt) coordinate (C2);	
		\draw (A1)--(C1); \draw (B2)--(C2); \draw (A2)--(B1); }
}

\allowdisplaybreaks
 
\begin{document}
	
	\title{Quantum crystals, Kagome lattice, and plane partitions\\ fermion-boson duality}
	
	
	\vskip 1 cm
	
	 \author{Thiago Araujo}
	 \email{thiago@itp.unibe.ch}
	 \affiliation{Albert Einstein Center for Fundamental Physics, Institute for Theoretical Physics, University of Bern, 
		Sidlerstrasse 5, ch-3012, Bern, Switzerland}
	 \author{Domenico Orlando}
	 \email{domenico.orlando@to.infn.it}
	 \affiliation{Albert Einstein Center for Fundamental Physics, Institute for Theoretical Physics, University of Bern, 
		Sidlerstrasse 5, ch-3012, Bern, Switzerland}
		\affiliation{INFN, sezione di Torino and Arnold-Regge Center, via Pietro Giuria 1, 10125 Torino, Italy}
	 \author{Susanne Reffert}
	 \email{sreffert@itp.unibe.ch}
	 \affiliation{Albert Einstein Center for Fundamental Physics, Institute for Theoretical Physics, University of Bern, 
		Sidlerstrasse 5, ch-3012, Bern, Switzerland}
	
	\vskip 1 cm
	

	\begin{abstract}
	\noindent
	In this work we study quantum crystal melting in three space dimensions. 
	Using an equivalent description in terms of dimers in a hexagonal lattice, we recast the 
	crystal melting Hamiltonian as an occupancy problem in a Kagome lattice. The Hilbert space is spanned by states labeled by
	plane partitions, and writing them as a product of interlaced integer partitions, we define a fermion-boson 
	duality for plane partitions. Finally, based upon the latter result we conjecture that the growth operators 
	for the quantum Hamiltonian can be represented in terms of the affine Yangian \({\cal Y}[\widehat{\mathfrak{gl}}(1)]\). 
	\end{abstract}
	
	\maketitle
	
	\setcounter{equation}{0}


	\section{Introduction} 	
	
	Random partitions appear in many contexts in mathematics and physics. This omnipresence is partly explained by the fact that they are part of the core of number theory, the queen of mathematics~\cite{von1856gauss}. It is nonetheless remarkable that they appear in a variety of distinct problems -- such as the Seiberg--Witten theory, 
	integrable systems, black holes, string theory and others~\cite{Okounkov2003,Okounkov:2003sp,Iqbal:2003ds,Nekrasov:2003rj,Heckman:2006sk}.
	
	In two dimensions, random partitions can be successfully realized in terms of fermionic 
	operators living on a chain where particles and holes are labeled by half-integers along the real line~\cite{Jimbo:1983if,Babelon2003,Okounkov:2003sp,Dijkgraaf:2008ua}. In this formalism, the empty partition is equivalent to a 
	configuration where all negative holes (or better yet, holes on negative positions) are occupied and all 
	positive holes are vacant.
	\begin{figure}[h]
		\centering
		\includegraphics[width=6.0cm]{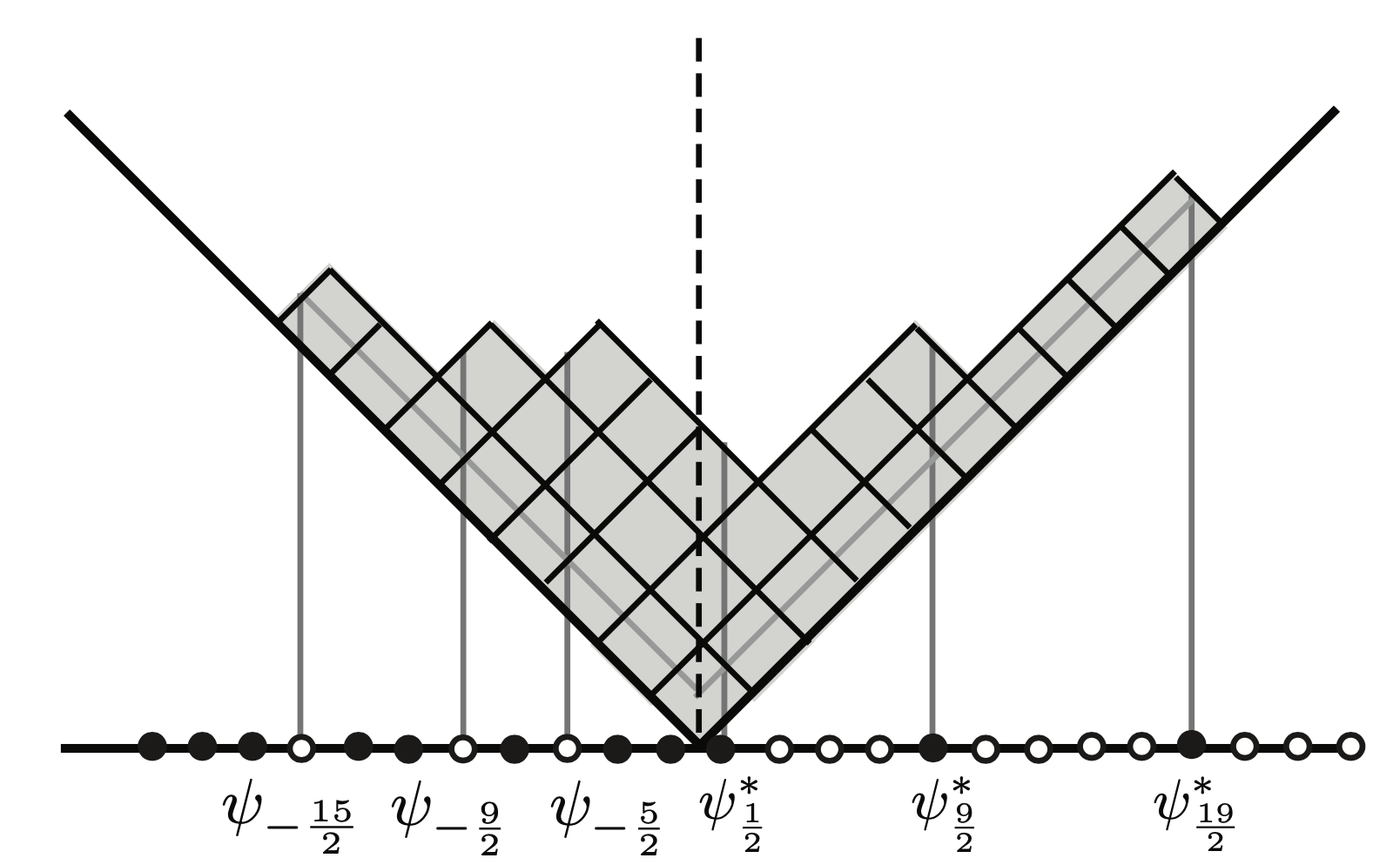}
		\caption{Fermionic fields and Young diagram relation.}
		\label{fig:intpart}
	\end{figure}
	The creation and annihilation of squares are defined as holes and particles hopping as in figure~(\ref{fig:intpart}). 
	
	The description above has been used in~\cite{Dijkgraaf:2008ua,Orlando:2009kd} to study the
	two-dimensional version of quantum crystal melting. Quite remarkably, with a Jordan-Wigner transformation, 
	the crystal melting Hamiltonian is shown to be equivalent to the XXZ spin chains with kink boundary conditions, and it 
	automatically implies the integrability of this problem.
	
\begin{figure}[h]
		\centering
		\includegraphics[width=7.0cm]{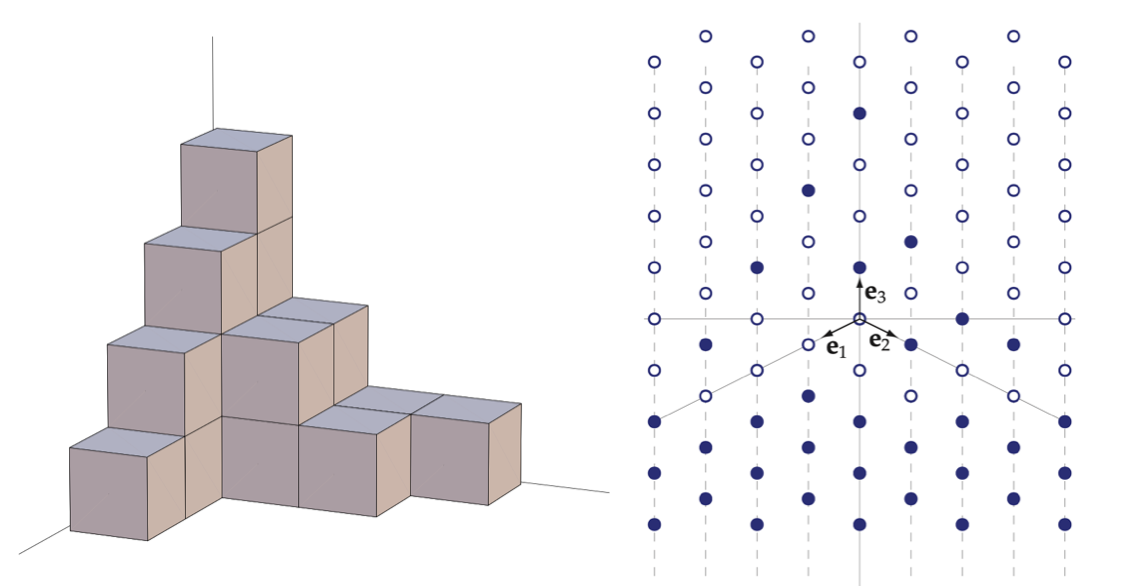}
		\caption{3D Young diagram and its fermions lattice}
		\label{fig:3dpart}
	\end{figure}

	In this work, we focus on the three-dimensional version of the quantum crystal melting, which can be rephrased in terms of 
	random plane partitions, and it has been partly addressed in~\cite{Dijkgraaf:2008ua}. Plane partitions 
	can be written in terms of interlaced XXZ diagrams or as lattice fermions, see figure~(\ref{fig:3dpart}).
	The empty partition is a configuration of the hexagonal lattice where all positions below the diagonals are occupied. 
	Additionally, based on the one- and two-dimensional cases, a conjecture for the mass gap has been made,
	and further numerical analysis has been performed.
	
	We present a number of equivalent formulations of the 3D quantum crystal melting problem, such as in terms of particle-hole
	hopping, a fermionic description on a Kagome lattice, and a tensor product representation.
	We are eventually lead to a fermion-boson duality, and to a conjecture for an underlying Yangian symmetry.
	We expect this structure to be instrumental in uncovering a possible hidden integrable structure, suggested by the results in lower dimensions.
	
	\medskip
	The plan of this note is as follows. We start with a short review in Section~\ref{sec:plane_hex}, where we see how plane partitions are equivalently written in 
	terms of dimers in a hexagonal lattice.

	As we said before, we want to find an alternative particle-hole dynamics which is equivalent to the plane 
	partition growth. More explicitly, we search for a statistical system with states labeled by 
	plane partitions. That is the problem we address in Section~\ref{sec:part-hole}, where we discuss a construction in terms of 
	particle hopping in a hexagonal lattice. This formalism is considerably simplified 
	if we use a dual description which turns out to be the Kagome lattice. We write the new Hamiltonian, 
	and we provide additional details of it in Section~\ref{sec:kagome}.
	
	Having discussed characterizations of the crystal melting as occupation problems, we try to 
	understand other fundamental aspects of the system, in particular, its Hilbert space. In Section~\ref{sec:tensorProd}, 
	we write the plane partition states in terms of interlaced integer partitions, and we 
	use this fact to show that the growth operators have nontrivial coproducts. In addition, we 
	define a novel fermion-boson correspondence for plane partitions states in Section~\ref{sec:fb-corr}. 
	Finally, based on the action of the growth operators on the bosonized states, 
	we conjecture that this Hamiltonian can represented in terms of the affine Yangian of \(\widehat{\mathfrak{gl}}(1)\). 
	We give some concluding remarks in Section~\ref{sec:conclusions}. In Appendix~\ref{app:corres}, we give present some necessary details on the 2D fermion-boson duality, in Appendix~\ref{app:kagome} we argue the equivalence of the Hamiltonians in the plane-partition language and in the Kagome picture. In Appendix~\ref{app:basis}, we detail basis transformations for various states and in Appendix~\ref{app:amplitudes}, we give amplitudes for transitions between various states.
	

	\section{Plane partitions \& Hexagonal Dimers}\label{sec:plane_hex}
	
\begin{figure}[h]
	\centering
	\includegraphics[width=4.0cm]{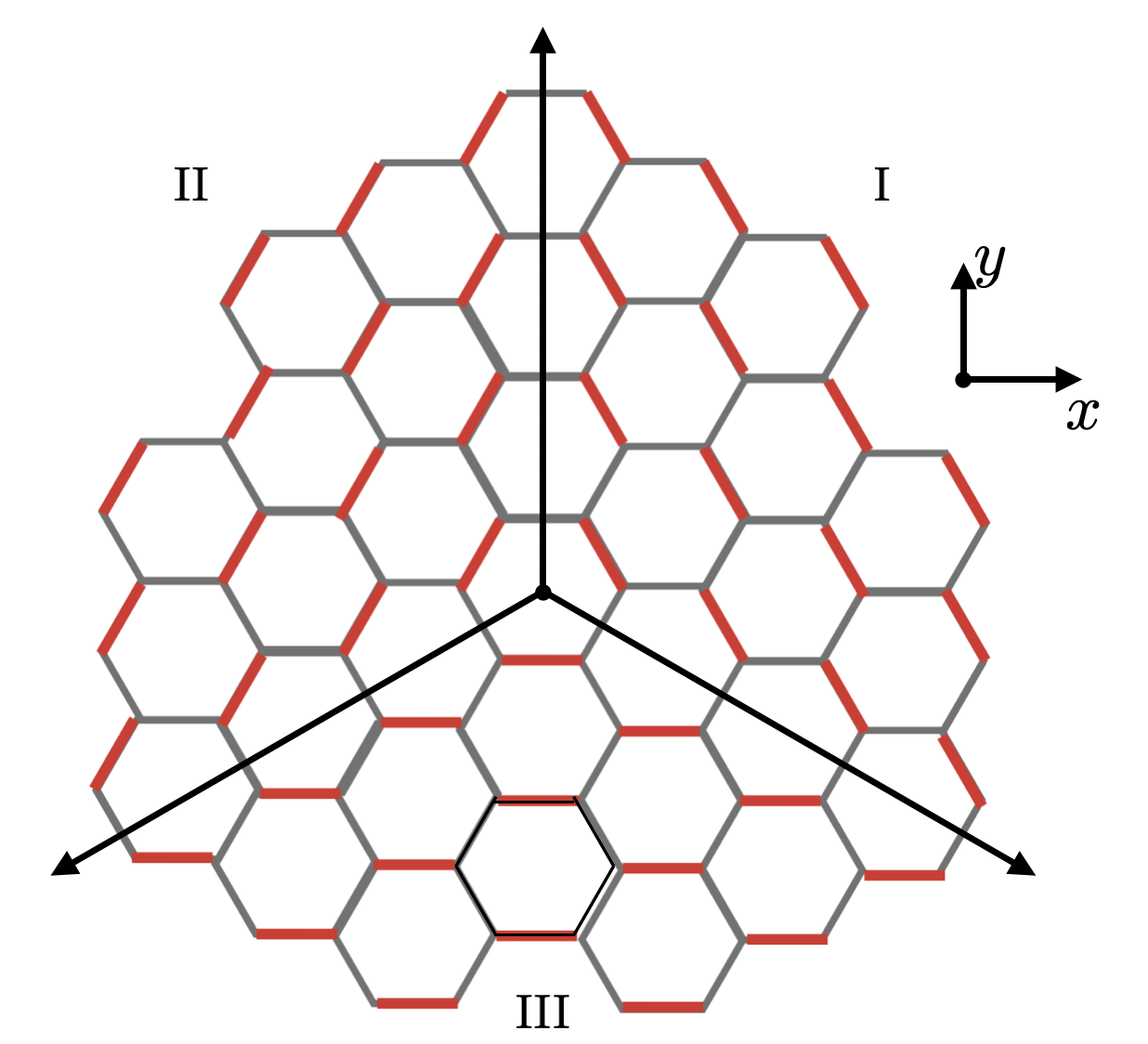}
	\caption{Empty partition}
	\label{fig:empty}
	\end{figure}

	We start by describing the well known correspondence between the crystal melting problem and dimers in a 
	hexagonal lattice, see~\cite{Moessner2008,Kenyon2008,Dijkgraaf:2008ua,Dijkgraaf:2007yr} and references therein. 
	The empty partition (vacuum) is equivalent to an infinity perfect matching with different staggered boundary 
	conditions in the regions \(\textrm{I}\), \(\textrm{II}\) and \(\textrm{III}\), see figure~(\ref{fig:empty}), defined as
	\be 
	\label{eq:regions}
	\begin{split}
	& \vartheta \in \textrm{I}=( - \pi/6,\pi/2 ),\\
	& \vartheta \in \textrm{II}=( \pi/2,7\pi/6),\\
	& \vartheta \in \textrm{III}=( - 5\pi/6,-\pi/6 ).
	\end{split}
	\ee	
	
	In the vacuum configuration, the origin is defined as the only plaquette 
	where the three phases, or staggered directions, meet. 
	We say that a plaquette with the three phases is a \emph{flippable plaquette}.
	
\begin{figure}[h]
		\centering
		\includegraphics[width=4.0cm]{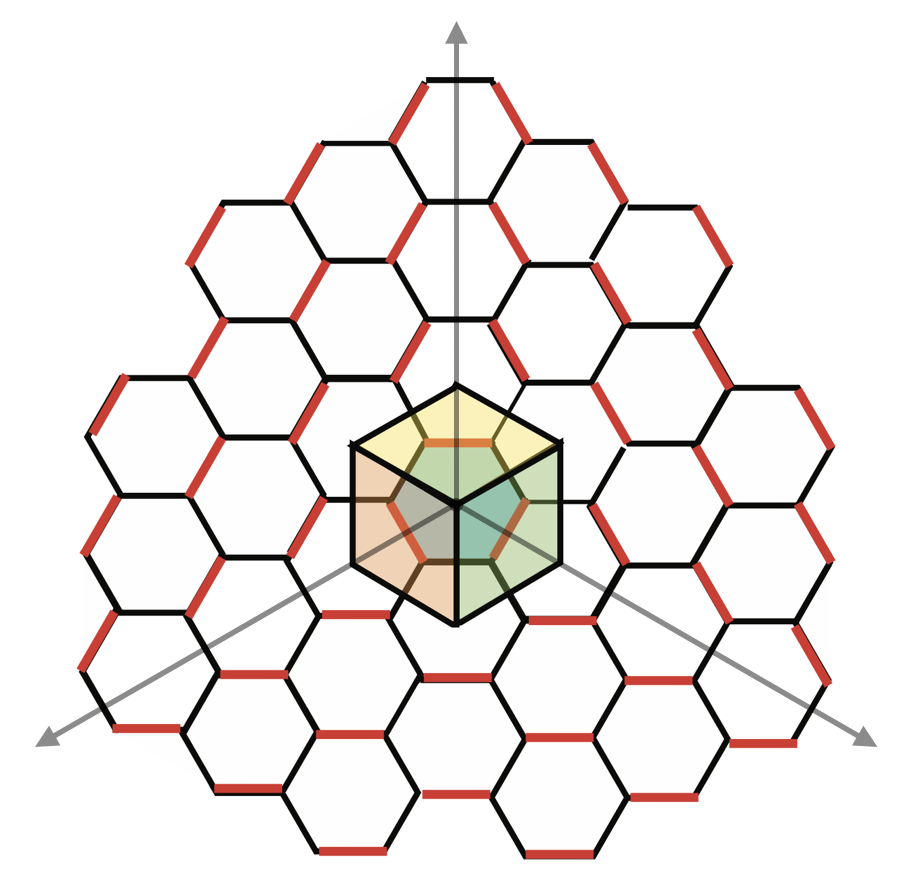}
		\caption{1 Box configuration}
		\label{fig:1box}
	\end{figure}

	Adding a box to the vacuum corresponds to the \emph{positive} flip \((\spp) \mapsto (\blo)\), as seen in figure~(\ref{fig:1box}),
	and as a matter of fact, there is only one flippable plaquette in the vacuum. Additionally, simple inspection of figure~(\ref{fig:1box})
	reveals that there are now three plaquettes where we can perform a positive flip, and just one plaquette where the
	\emph{negative} flip \((\blo) \mapsto (\spp)\) can be performed. 
	
	It is easy to see that this logic remains the same for general configurations, that is, positive and negative flips correspond
	to adding and removing boxes in a given plane partition state. Furthermore, for any lattice configuration, there 
	is only a finite number of flippable plaquettes, and from the plane partition perspective it obviously means that there is only a finite
	number of boxes that can be consistently removed, or available places where we can add a box.  
	
	We want to study the crystal melting Hamiltonian defined in~\cite{Dijkgraaf:2008ua}
	\bse
	\be
	\begin{split}
	H = & -J \sum \left( |\blo\rangle \langle \spp |  + |\spp\rangle \langle \blo |\right) \\ 
	\hspace{1.0cm} & + V \sum\left( \sqrt{q} |\spp\rangle \langle \spp | + \frac{1}{\sqrt{q}}|\blo\rangle \langle \blo | \right)\; ,
	\end{split}
	\ee
	that we conveniently write as
	\be
	\label{eq:Hamiltonian}
	\begin{split}
		H = & - J \left( \Gamma + \Pi \right) + V\left( 
		\sqrt{q} \; {\cal O}_+(\Pi, \Gamma) +\frac{1}{\sqrt{q}} {\cal O}_-(\Pi, \Gamma)
		\right)\; .
	\end{split}
	\ee
	\ese
	Using \(|\blo\rangle^\dagger =\langle \blo| \) and \( |\spp\rangle^\dagger = \langle \spp|\), one can easily see that the Hamiltonian is Hermitian.
	The two terms multiplied by $J$ are kinetic and describe the two flips 
	previously mentioned. More specifically, the operator \(\Gamma\) (\(\Pi\)) adds (removes) boxes in all possible ways. 
	Moreover, the potential terms \({\cal O}_+(\Pi, \Gamma) \) and \({\cal O}_-(\Pi, \Gamma)\) count the 
	number of positive and negative flippable plaquettes, respectively. In other words, they determine 
	the number of places where we can add a box, and the number of boxes that can be removed 
	respecting the plane partition rules.

   This system has a special point for \(J = V\), the so-called Rokhsar-Kivelson point~\cite{Rokhsar:1988zz,Henley2003}, where it is natural to understand the model as the stochastic quantization of the classical counting of plane partitions~\cite{Dijkgraaf:2009gr}. In the continuum limit this corresponds to a quantum critical point~\cite{Ardonne:2003wa}.
   At this point we expect the system to have special symmetry properties and we will focus on this case.
	Some progress in the understanding of this system has been achieved in~\cite{Dijkgraaf:2008ua}, where it was shown that the ground state of the system is given by a sum over all plane partitions, where a partition of the integer \(k \) is weighted by \(q^{k/2 }\).
   If follows that the norm of the ground state is given by
	\be 
	\label{eq:part.function}
	Z\equiv \langle \textrm{ground}|\textrm{ground}\rangle = \prod_{n\geq 1} \frac{1}{(1-q^n)^n}\; ,
	\ee
	which we recognize as the famous {MacMahon function}: the generating function for the 
	number \(pl(k)\) of plane partitions 
	\be 
	Z = \sum_{k=0}^\infty pl(k) q^k\; .
	\ee 
	In the following sections, we want to represent the Hamiltonian~(\ref{eq:Hamiltonian}) in terms of a statistical mechanics systems
	of particles and holes, and we start investigating its integrability properties. More specifically, we define a new form of
	fermion-boson correspondence and we find connections between this problem and the affine Yangians of \(\widehat{\mathfrak{gl}}(1)\).


	\section{Particle-hole bound states}\label{sec:part-hole}
	
	In this section we describe the dynamics of the crystal melting problem in terms of particle hopping in two Bravais lattices (\ref{eq:Hamiltonian}). In the next section we build a Hamiltonian for this problem using the dual description of this 
	particle-hole system. 
	
	First we need to remember that the hexagons above do not define a Bravais lattice, but we can write it as
	a superposition of two triangular Bravais lattices, say \texttt{H} whose points are labeled by \(\circ\), 
	and \texttt{P}, labeled by \(\bullet\). We call them \emph{Hole} and \emph{Particle} sublattices, respectively.
	
\begin{figure}[h]
		\centering
		\includegraphics[width=3.0cm]{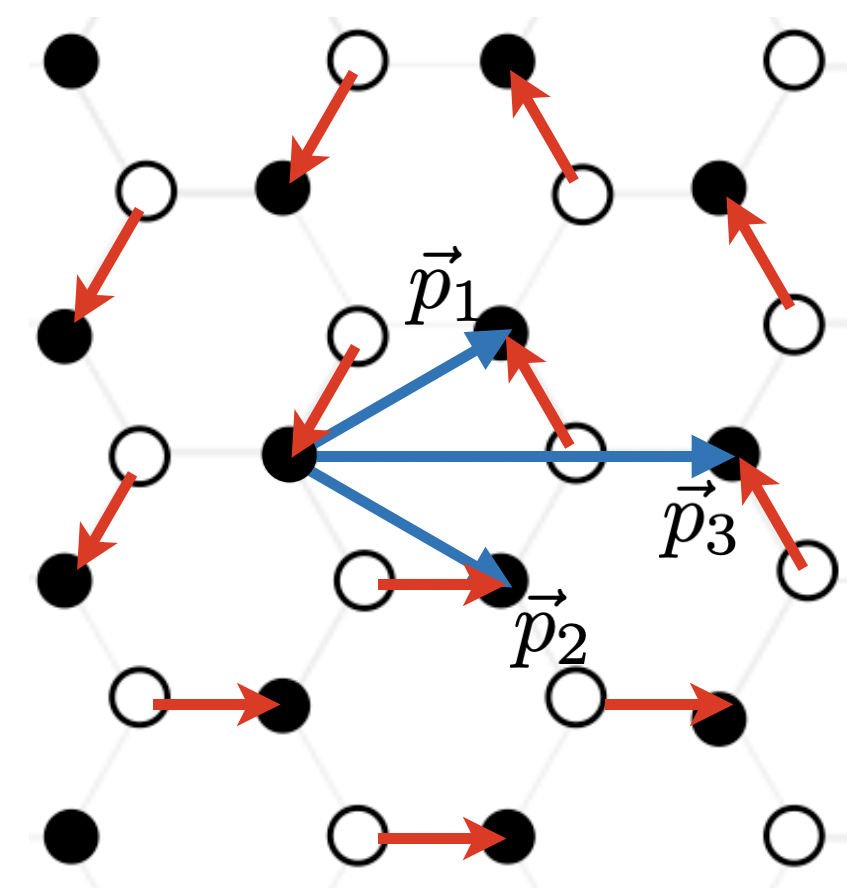}
		\caption{Sublattices \texttt{H} (\(\circ\)) and \texttt{P} (\(\bullet\))}
		\label{fig:bravais}
	\end{figure}

	The position vectors in the \texttt{P} sublattice are
	\be 
	\vec{p}_1 = \frac{a}{2}(\sqrt{3}, 1), \qquad \vec{p}_2 = \frac{a}{2}(\sqrt{3}, -1),
	\ee
	where \(a\) is the lattice spacing, that can, without loss of generality, be set to \(a=1\), see details in~\cite{Abergel_2010}. 
	In figure~(\ref{fig:bravais}) we have also
	given the linear combination \(\vec{p}_3 = \vec{p}_1 + \vec{p}_2\). 
	Finally, the dimer between two lattice points is represented by the positively oriented 
	bound states \(\circ \to \bullet \).
	
	The fermionic operators \(\psi(\vec{x})\) and \(\psi^\ast(\vec{x})\) in an arbitrary lattice 
	point \(\vec{x}\) satisfy the anticommutation relations 
	\be 
	\begin{split}
	& \{\psi(\vec{x}), \psi(\vec{x}') \} = 0, \\
	& \{\psi^\ast(\vec{x}), \psi^\ast(\vec{x}') \} =0, \\
	& \{\psi^\ast(\vec{x}), \psi(\vec{x}') \} =\delta_{\vec{x}, \vec{x}'}\; .
	\end{split}
	\ee
	The \emph{bottom} of the Dirac sea is defined by the condition
	\be 
	\psi(\vec{x}) |0\rangle = 0\; , \quad \forall \ \vec{x} \in \texttt{H}, \texttt{P} \; ,
	\ee
	and the empty partition (vacuum) by
	\be 
	|\bm{\emptyset}\rangle =
	\prod_{j\in \mathbb{Z}}\prod_{k\in \mathbb{Z}} \psi^\ast(j \vec{p}_1+ k \vec{p}_2) |0\rangle\; .
	\ee
	With these definitions, it is straightforward to see that all points in the sublattice \texttt{P} 
	are occupied. It should be implicit (since it is a visual aid) that the holes and 
	particles are bonded positively, \(\circ \to \bullet \).  A flip in the hexagon (creation or annihilation of a box) is simply a 
	hole-particle hopping which preserves the positive orientation of the bound state. 
	
\begin{figure}[h]
		\centering
		\includegraphics[width=7.0cm]{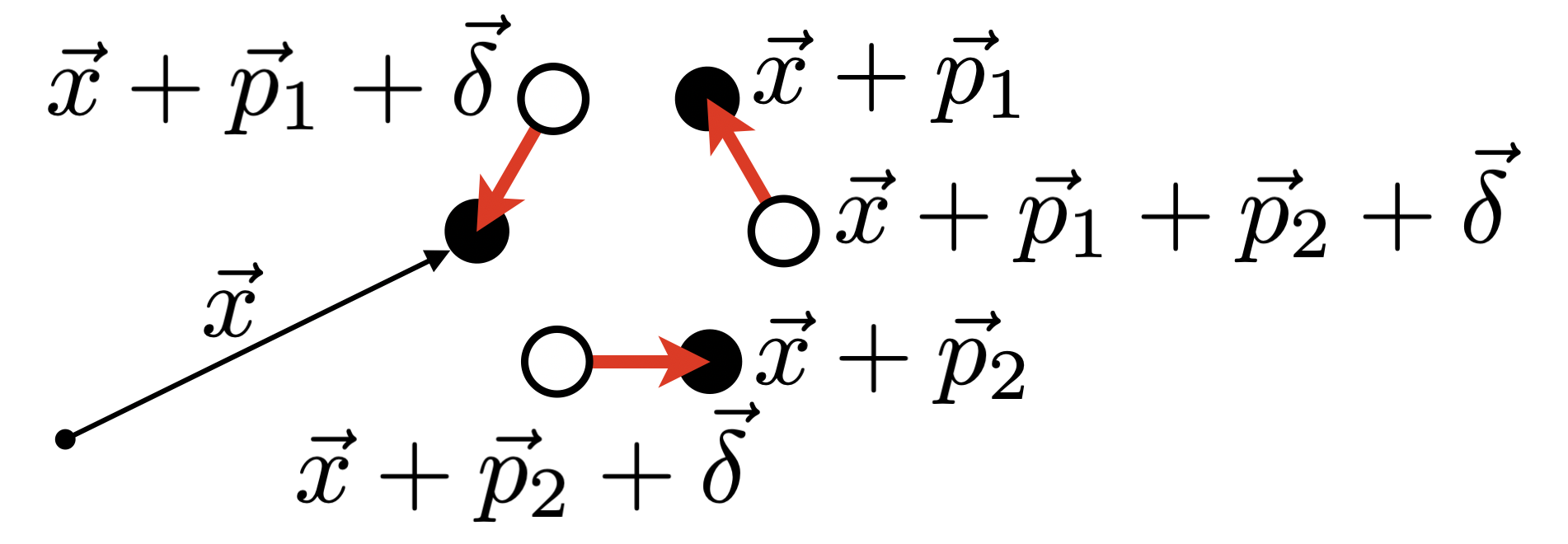}
		\caption{Generic points in the lattice.}
		\label{fig:vector}
	\end{figure}

	Furthermore, we set the origin at \((j,k)=\vec{0}\), and we can identify a generic point
	\(\vec{x}=j \vec{p}_1 + k \vec{p}_2\), \((j,k) \in \mathbb{Z}^2\) as in figure~(\ref{fig:vector}).
	
	Points in the Hole sublattice \texttt{H} can be identified by linear combinations of \((\vec{p}_1, \vec{p}_2, \vec{p}_3)\)
	plus the additional offsets 
	\be 
	\vec{\delta}_\pm = \frac{a}{2}\left( \frac{1}{\sqrt{3}}, \pm 1\right)\; , \quad 
	\vec{\delta} = a\left( -\frac{1}{\sqrt{3}}, 0\right)\; .
	\ee
	Putting all these facts together, we can try to build a model where the flips in the hexagonal dimers correspond to simultaneous 
	hoppings in the two sublattices.  
	
\begin{figure}[h]
		\centering
		\includegraphics[width=5.0cm]{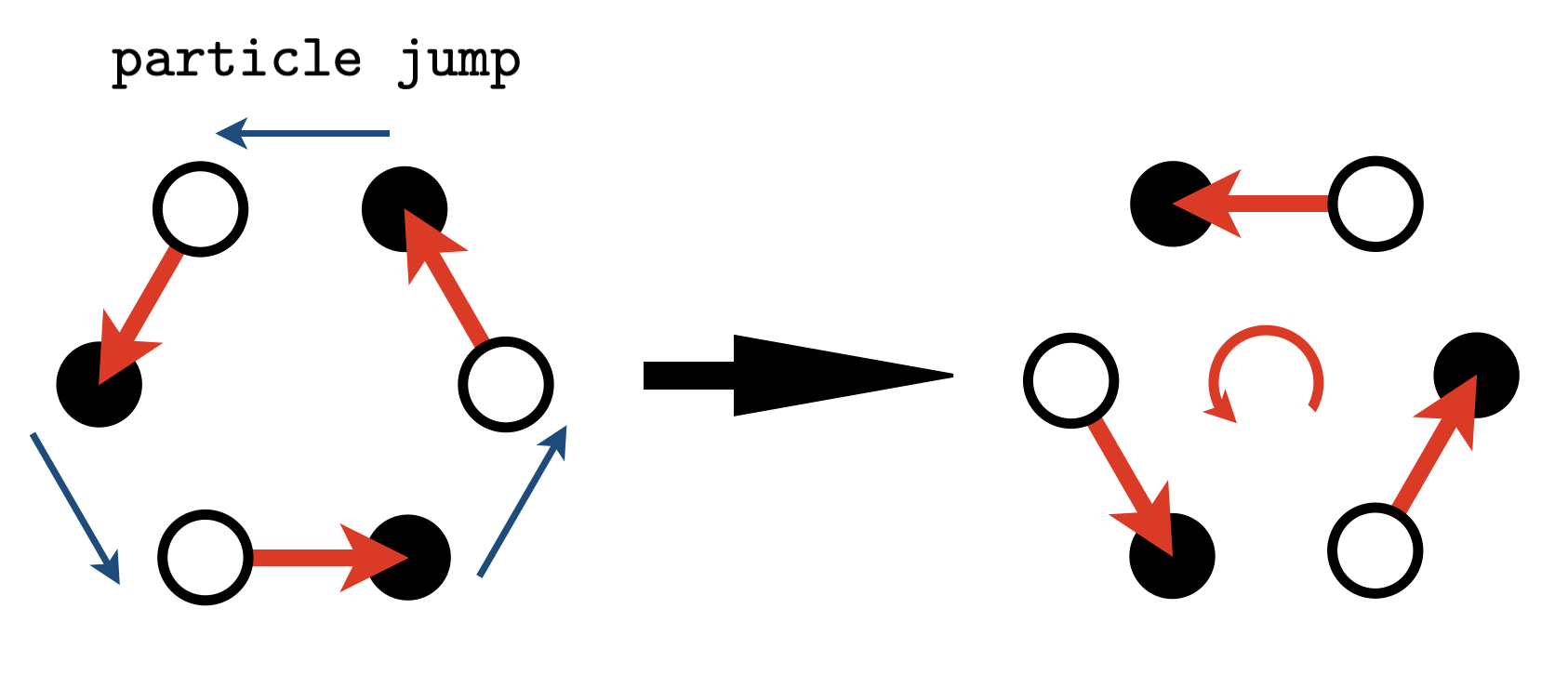}
		\caption{Counterclockwise 3-hopping}
		\label{fig:chopping}
	\end{figure}

	The creation of one box is described as a counterclockwise rotation as in figure~(\ref{fig:chopping}), and it is equivalent to 
	three simultaneous hoppings, a \emph{3-hopping}, along the positively oriented (the red arrow) hole-particle bound states,
	\bse
	\be 
	\label{eq:box-creation}
	\begin{split}
	\ytableausetup{centertableaux, smalltableaux}
	| \ydiagram[*(lightgray)]{1} \rangle = & \psi^\ast(\vec{p}_1+\vec{\delta})\psi (\vec{p}_1) \\
	& \psi^\ast(\vec{p}_2+\vec{\delta})\psi(\vec{0})
	\psi^\ast(\vec{p}_1 + \vec{p}_2 +\vec{\delta})\psi (\vec{p}_2) |\bm{\emptyset}\rangle. 
	\end{split}
	\ee
	
\begin{figure}[h]
		\centering
		\includegraphics[width=5.0cm]{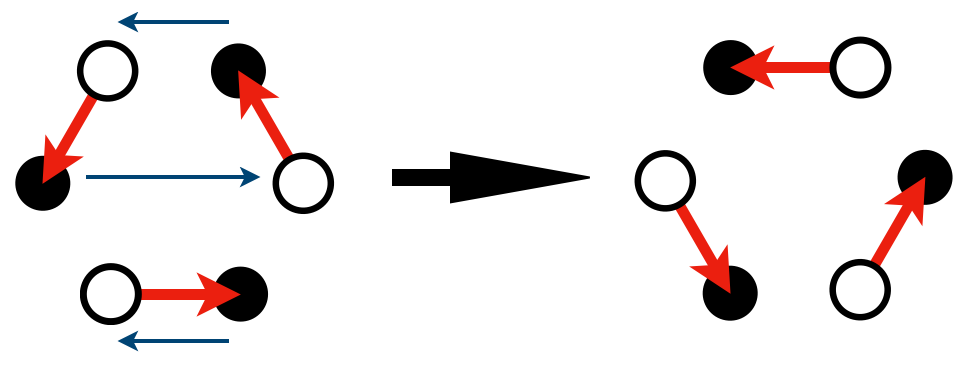}
		\caption{Horizontal 3-hopping}
		\label{fig:hhopping}
	\end{figure}

	This point of view describes the box creation in terms of next-neighbor interactions, but it may be computationally useful 
	to describe the box creation in terms of the \emph{horizontal} 3-hopping that we define as in figure~(\ref{fig:hhopping}). 
	
	In other words, we reshuffle the expression~(\ref{eq:box-creation})
	\be 
	\begin{split}
	\ytableausetup{centertableaux, smalltableaux}
	| \ydiagram[*(lightgray)]{1} \rangle = & \psi^\ast(\vec{p}_2+\vec{\delta})\psi(\vec{p}_2)
	\psi^\ast(\vec{p}_1 + \vec{p}_2 +\vec{\delta})\psi (\vec{0})\\ 
	& \psi^\ast(\vec{p}_1+\vec{\delta})\psi (\vec{p}_1)|\bm{\emptyset}\rangle \; .
	\end{split}
	\ee
	\ese
	
	The 3-hopping above describes the creation and annihilation of a state corresponding to a partition with one box. 
	For configurations with more boxes, some of the relevant bound states are edges of two hexagons. 
	In this case, we would need to consider a horizontal 2-hopping where there are two free bound states, 
	and the usual hopping when there is just one free bound state.

	In other words, we write the 3 configurations with 2 boxes as
	\bse
	\be 
	\begin{split}
	\ytableausetup{centertableaux, smalltableaux}
	\left| \ydiagram[*(lightgray)]{1,1}\right\rangle = &
	\psi^\ast(2\vec{p}_1 +\vec{\delta})\psi (\vec{p}_1-\vec{p}_2)\\
	&\psi^\ast(2 \vec{p}_1 - \vec{p}_2+ \vec{\delta})\psi(2 \vec{p}_1 - \vec{p}_2)
	| \ydiagram[*(lightgray)]{1}\rangle\; ,
	\end{split}
	\ee
	\be
	\begin{split}
	\ytableausetup{centertableaux, smalltableaux}
	 \left| \ydiagram[*(lightgray)]{2}\right\rangle =  &
	\psi^\ast(2\vec{p}_2 +\vec{\delta})\psi (\vec{p}_1+\vec{p}_2)\\
	& \psi^\ast(\vec{p}_1+  2\vec{p}_2+\vec{\delta})\psi (2\vec{p}_2)
	| \ydiagram[*(lightgray)]{1}\rangle ,
	\end{split}
	\ee
	and 
	\be 
	\begin{split}
	\ytableausetup{centertableaux, smalltableaux}
	| \ytableaushort[*(lightgray)]{2 }\rangle = &
	\psi^\ast(-\vec{p}_1+\vec{p}_2+\vec{\delta})\psi(-\vec{p}_1)\\
	& \psi^\ast(\vec{\delta}) \psi (-\vec{p}_1+\vec{p}_2)
	| \ydiagram[*(lightgray)]{1}\rangle\; .
	\end{split}
	\ee
	\ese
	
	This description is rather involved, and somewhat contrived, for states with more than one box. Fortunately, in the next section
	we show that there exists a more natural description when we assume that for each dimer, there is a dual description 
	in which this extended object is pointlike.

	
	\section{Dual Kagome lattice}\label{sec:kagome}
	
\begin{figure}[h]
		\centering
		\includegraphics[width=3.0cm]{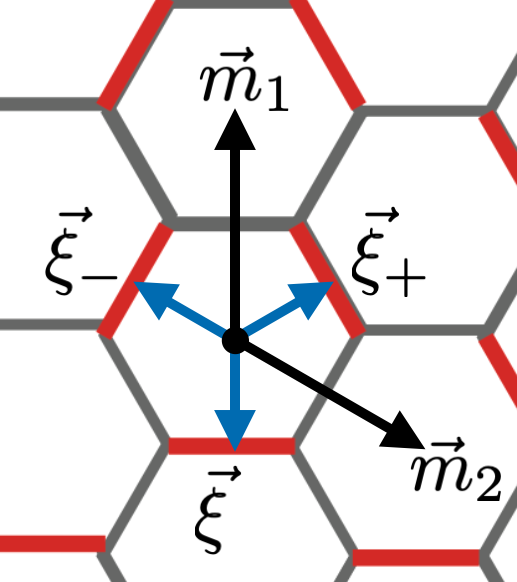}
		\caption{Dual basis}
		\label{fig:dbasis}
	\end{figure}

	In order to define this dual description, let us start with the basis
	\be 
	\vec{m}_1 = a(0,1)\; , \quad 
	\vec{m}_2 = \frac{a}{2}(\sqrt{3}, -1 )\; ,
	\ee
	connecting the barycenters of hexagons, as in figure~(\ref{fig:dbasis}).
	Furthermore, the offset vectors from the center of a hexagon to its edges 
	are
	\be 
	\vec{\xi}_\pm = \frac{a}{4}(\pm \sqrt{3}, 1 )\; , \quad
	\vec{\xi} = \frac{a}{2}(0,-1)\; .
	\ee
	
\begin{figure}[h]
		\centering
		\includegraphics[width=2.5cm]{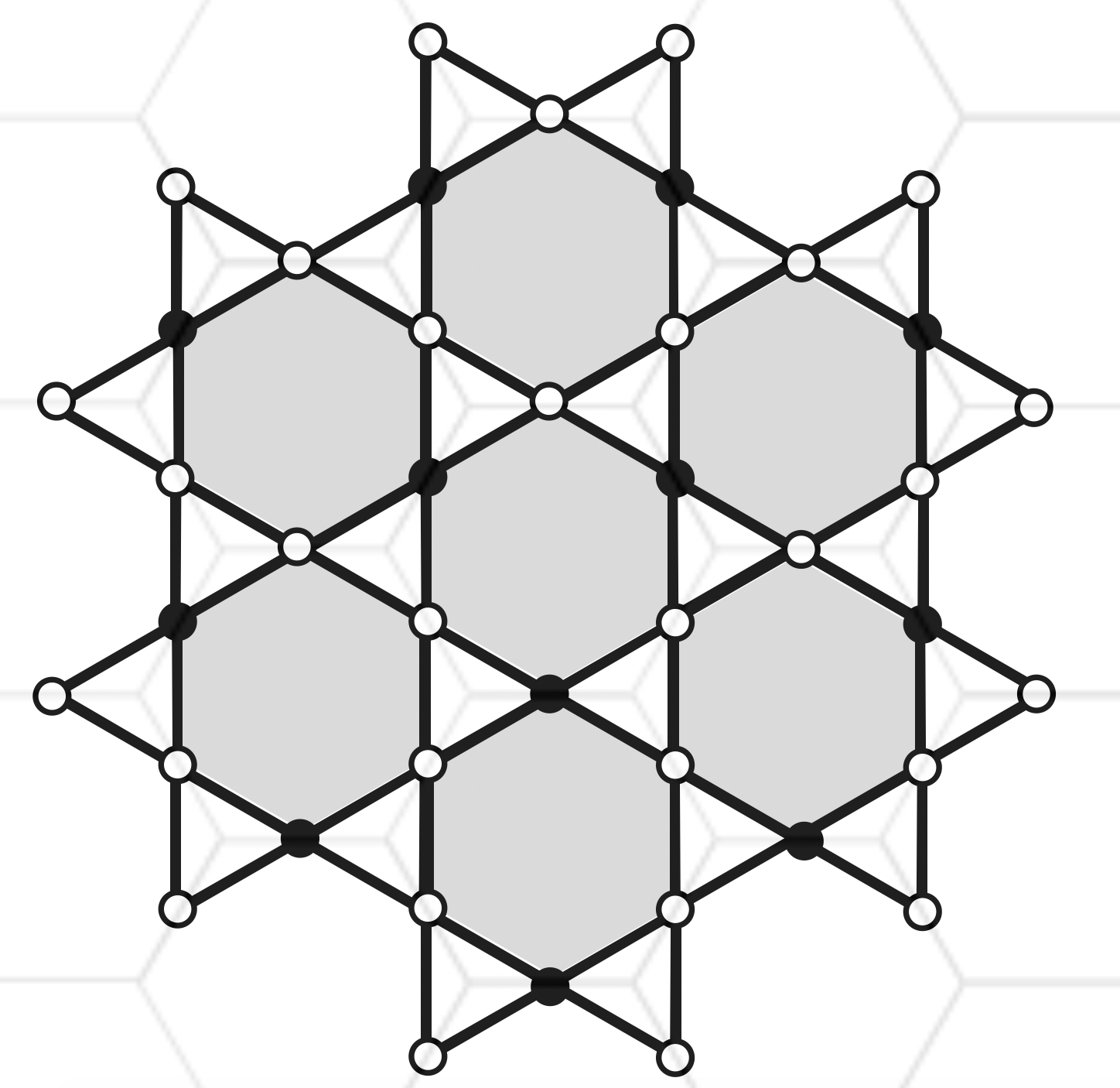}
		\caption{Empty configuration in the Kagome lattice}
		\label{fig:tessel}
	\end{figure}

	Given a hexagon in the lattice, 
	let us associate a \emph{dual} fermion (created by \(\tilde{\psi}^\ast\)) to the midpoint of each edge containing a dimer and 
	a \emph{dual} hole (where the fermion is annihilated by \(\tilde{\psi}\)) to the midpoint of each empty edge. Connecting these midpoints generates another lattice made of 
	hexagons and triangles, which is known as \emph{Archimedean tessellation} \(\{3,6,3,6\}\)~\cite{lang2017}, or \emph{Kagome lattice}. The empty configuration corresponds to figure~(\ref{fig:tessel}).
	
\begin{figure}[h]
		\centering	
		\includegraphics[width=2.5cm]{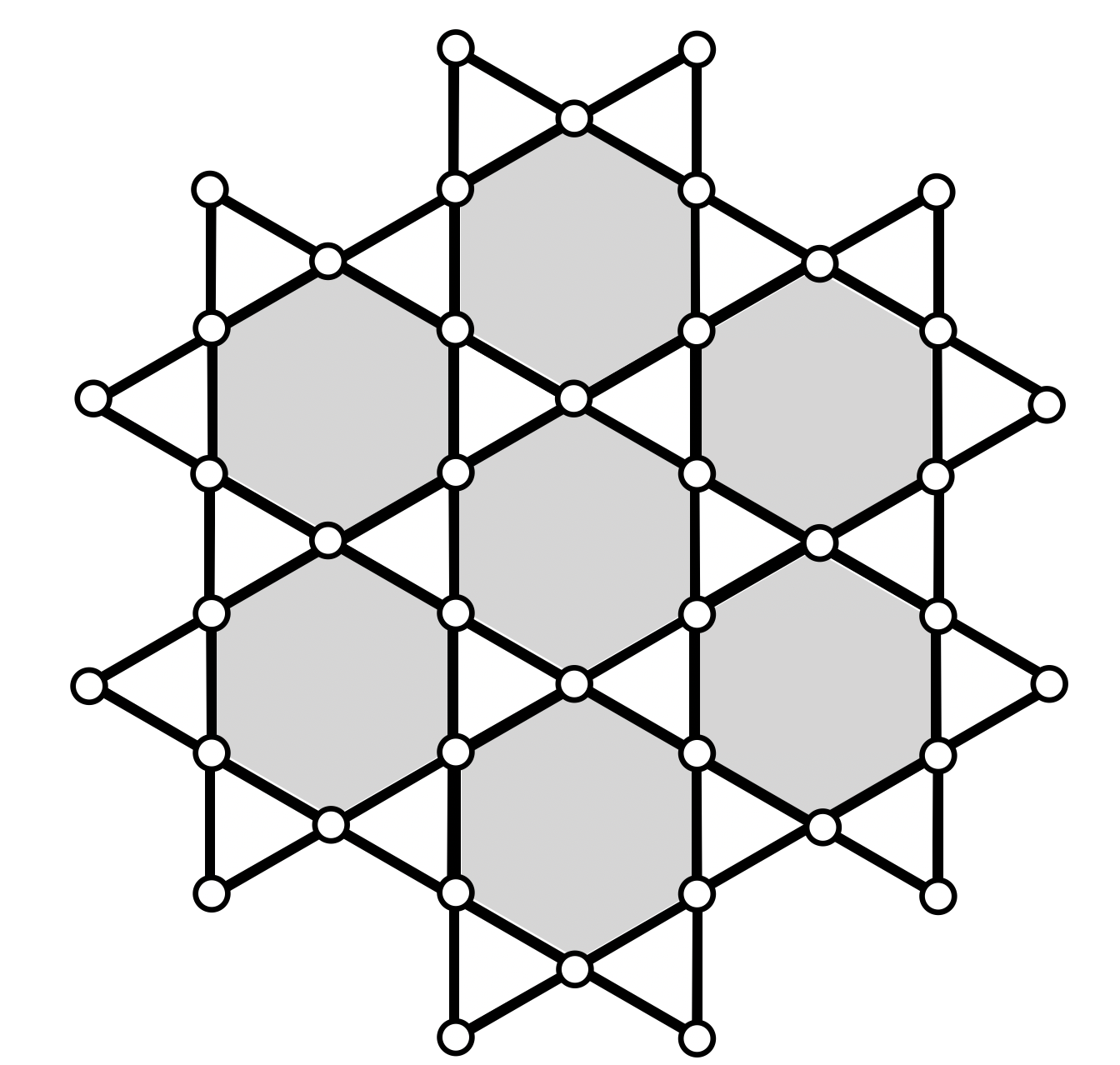}
		\caption{Dual Dirac sea.}
		\label{fig:hex.trian}
	\end{figure}

	In terms of the dual fermions, the empty configuration is given by 
	\bse
	\be 
	|\bm{\emptyset}\rangle = 
	\prod_{j\in \mathbb{Z}}\prod_{k\in \mathbb{Z}} 
	\tilde{\psi}^\ast(j \vec{m}_1 + k \vec{m}_2 + \Delta_{j,k}) |\tilde{0}\rangle\; ,
	\ee
	where \(|\tilde{0}\rangle\) is the Dirac sea in the Kagome lattice, which is defined by putting a dual hole at 
	the midpoint of each honeycomb edge, or equivalently, in the \(\{3,6,3,6\}\) tessellation, see figure~(\ref{fig:hex.trian}). 
	
	Moreover, we have defined
	\be 
	\begin{split}
	\Delta_{j,k} & = \vec{\xi}_+ \delta_{\textrm{I}}[\arg(j \vec{m}_1 + k \vec{m}_2)] + \\
	& \vec{\xi}_- \delta_{\textrm{II}}[\arg(j \vec{m}_1 + k \vec{m}_2)] 
	+ \vec{\xi} \delta_{\textrm{III}}[\arg(j \vec{m}_1 + k \vec{m}_2)]\; ,
	\end{split}
	\ee
	where
	\be 
	\delta_{\textrm{A}}[\arg(\vec{v})] = 
	\left\{
	\begin{array}{ll}
		0 & \text{if } \arg(\vec{v})\notin \textrm{A}\\
		1 & \text{if } \arg(\vec{v})\in \textrm{A}
	\end{array}
	\right. \; , \quad \textrm{A}=\textrm{I}, \textrm{II}, \textrm{III}\; ,
	\ee
	and the regions \textrm{I}, \textrm{II} and \textrm{III} have been defined in Eq.~(\ref{eq:regions}), see also 
	figure~(\ref{fig:empty}). Moreover, it is easy to write an explicit expression for \(\delta_{\textrm{A}}[\arg(\vec{v})]\) 
	as a difference of Heaviside functions.
	\ese
	
\begin{figure}[h]
		\centering	
		\includegraphics[width=2.5cm]{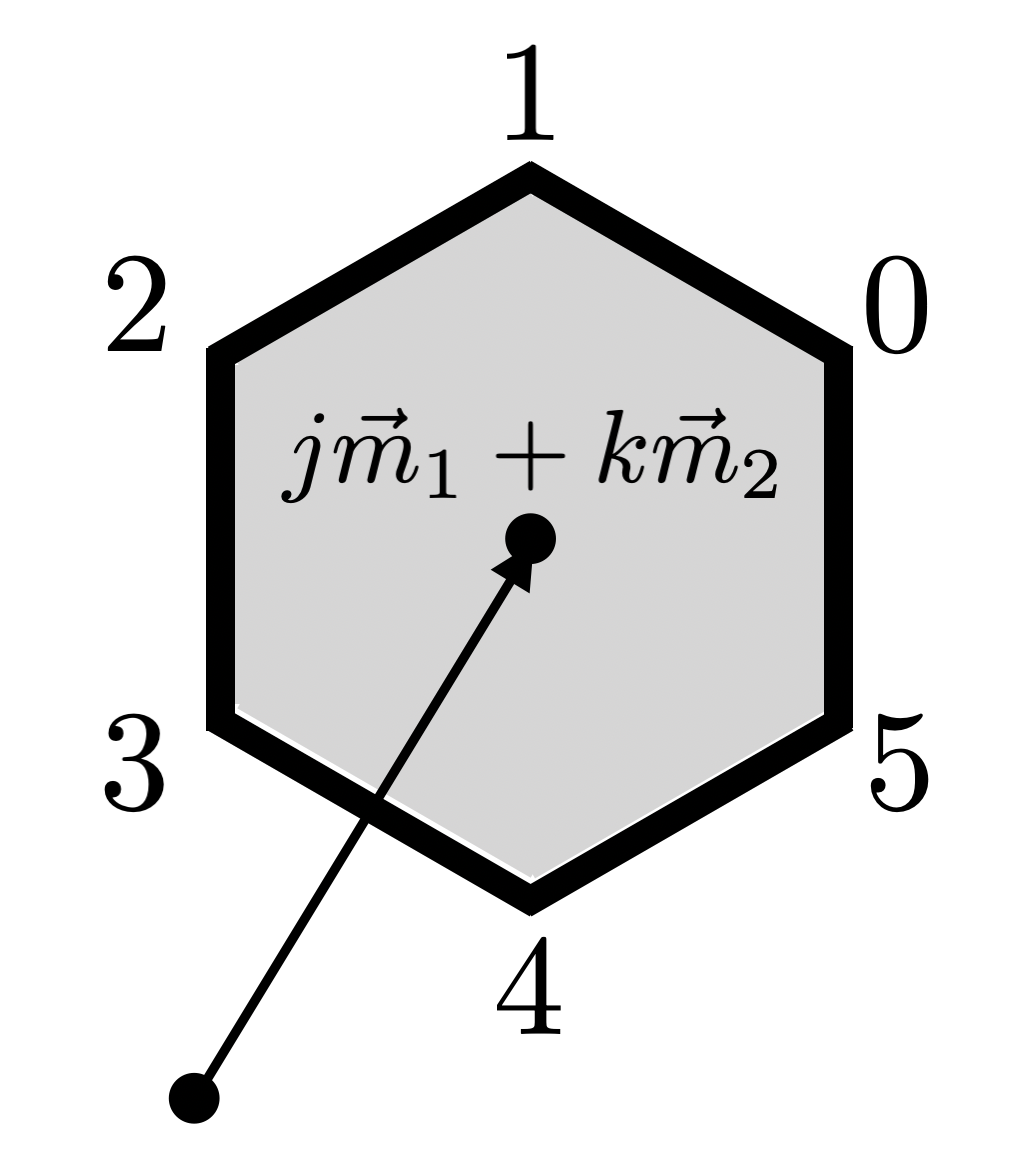}
		\caption{Vector position in the dual lattice}
		\label{fig:position}
	\end{figure}

	Using the dual lattice, the box creation and annihilation are given by the simultaneous action of three pairs of dual hole-particle pairs
	\(\tilde{\psi}^\ast\tilde{\psi}\) at neighboring vertices. The vectors from the origin to the vertices are given by
	\be 
	\begin{split}
	\vec{P}_{j,k}^{(s)} & = j \vec{m}_1 + k \vec{m}_2 \\ 
	& + \frac{a}{2}(\cos( (2 s+1)\pi/6),  \sin( (2 s+1)\pi/6) ),\\ 
	&s=0,1,2,3,4,5\; ,
	\end{split}
	\ee
	where we have introduced an index \(s\) as in figure~(\ref{fig:position}).
	
	Putting all these facts together, the 1-box configuration is given by
	\bse
	\be 
	\begin{split}
		\ytableausetup{centertableaux, smalltableaux}
		| \ydiagram[*(lightgray)]{1} \rangle = & \tilde{\psi}^\ast(\vec{P}_{0,0}^{(1)}) \tilde{\psi}(\vec{P}_{0,0}^{(0)})
		\tilde{\psi}^\ast(\vec{P}_{0,0}^{(3)}) \tilde{\psi}(\vec{P}_{0,0}^{(2)})\\ 
		& \tilde{\psi}^\ast(\vec{P}_{0,0}^{(5)}) \tilde{\psi}(\vec{P}_{0,0}^{(4)}) |\bm{\emptyset}\rangle \; 
	\end{split}
	\ee
	with inverse 
	\be 
	\begin{split}
		\ytableausetup{centertableaux, smalltableaux}
		| \bm{\emptyset} \rangle = & \tilde{\psi}^\ast(\vec{P}_{0,0}^{(0)}) \tilde{\psi}(\vec{P}_{0,0}^{(1)})
		\tilde{\psi}^\ast(\vec{P}_{0,0}^{(2)}) \tilde{\psi}(\vec{P}_{0,0}^{(3)}) \\ 
		& \tilde{\psi}^\ast(\vec{P}_{0,0}^{(4)}) \tilde{\psi}(\vec{P}_{0,0}^{(5)})
		| \ydiagram[*(lightgray)]{1} \rangle  \; .
	\end{split}
	\ee
	\ese
	
	Making the notation lighter with 
	\(\tilde{\psi}^{(s)}_{j,k}\equiv \tilde{\psi}(\vec{P}^{(s)}_{j,k})\), we have
	\bse
	\begin{align}
		\ytableausetup{centertableaux, smalltableaux}
		| \ydiagram[*(lightgray)]{1} \rangle & = \sum_{j,k \in \mathbb{Z}}\tilde{\psi}_{j,k}^{(1) \ast} \tilde{\psi}_{j,k}^{(0)}
		\tilde{\psi}_{j,k}^{(3) \ast} \tilde{\psi}_{j,k}^{(2)} \tilde{\psi}_{j,k}^{(5) \ast} \tilde{\psi}_{j,k}^{(4)} |\emptyset\rangle \\
		& \equiv \sum_{j,k \in \mathbb{Z}} \Gamma_{j,k} |\bm{\emptyset}\rangle\nonumber \\
		\ytableausetup{centertableaux, smalltableaux}
		|\bm{\emptyset}\rangle & = \sum_{j,k \in \mathbb{Z}} 
		\tilde{\psi}_{j,k}^{(0) \ast} \tilde{\psi}_{j,k}^{(1)}
		\tilde{\psi}_{j,k}^{(2) \ast} \tilde{\psi}_{j,k}^{(3)} \tilde{\psi}_{j,k}^{(4) \ast} \tilde{\psi}_{j,k}^{(5)} 
		| \ydiagram[*(lightgray)]{1} \rangle \\
		& \equiv \sum_{j,k \in \mathbb{Z}} \Pi_{j,k}  | \ydiagram[*(lightgray)]{1} \rangle\; .\nonumber
	\end{align}
	Observe that the action of the operator \(\Gamma_{j,k}\) on \(|\bm{\emptyset}\rangle\) for any position 
	\(j\neq 0\) and \(k\neq 0\) naturally vanishes because there is a mismatch of fermions and holes. 
	\ese	
	
	As the notation above suggests, we represent the operators \(\Gamma\) and \(\Pi\) in (\ref{eq:Hamiltonian}) 
	as
	\bse
	\begin{align}
	\Gamma & = \sum_{j,k} \Gamma_{j,k} = \sum_{j,k \in \mathbb{Z}}\tilde{\psi}_{j,k}^{(1) \ast} \tilde{\psi}_{j,k}^{(0)}
	\tilde{\psi}_{j,k}^{(3) \ast} \tilde{\psi}_{j,k}^{(2)} \tilde{\psi}_{j,k}^{(5) \ast} \tilde{\psi}_{j,k}^{(4)} \\
	\Pi & = \sum_{j,k} \Pi_{j,k} = \sum_{j,k \in \mathbb{Z}} 
	\tilde{\psi}_{j,k}^{(0) \ast} \tilde{\psi}_{j,k}^{(1)}
	\tilde{\psi}_{j,k}^{(2) \ast} \tilde{\psi}_{j,k}^{(3)} \tilde{\psi}_{j,k}^{(4) \ast} \tilde{\psi}_{j,k}^{(5)} \; .
	\end{align}
	\ese
	One may also calculate the commutators among the components
	\(\Gamma_{ij}\) and \(\Pi_{ij}\) of the growth operators. It is easy to see that the only nontrivial commutator 
	is \([\Gamma_{ij}, \Pi_{kl}]\). Although it does not give any new information, it is useful to have its explicit expression 
	since it is important in our calculations. It is straightforward (but laborious) to check the following relation:
	\begin{widetext}
		\be 
		\begin{split}
			\label{algebra}
			[\Gamma_{ij}, \Pi_{kl}] = \delta_{ik}\delta_{jl} & 
			\left( 
			\psi^{(\bar{5}4\bar{3}2\bar{1})}_{ij} \psi^{(1\bar{2}3\bar{4}5)}_{kl}
			-\psi^{(\bar{0})}_{kl} \psi^{(\bar{5}4\bar{3}2)}_{ij} \psi^{(\bar{2}3\bar{4}5)}_{kl} \psi^{(0)}_{ij}
			+ \psi^{(\bar{0}1)}_{kl} \psi^{(\bar{5}4\bar{3})}_{ij} \psi^{(3\bar{4}5)}_{kl} \psi^{(\bar{1}0)}_{ij}
			\right. \\
			& \left.\quad - \psi^{(\bar{0}1\bar{2})}_{kl} \psi^{(\bar{5}4)}_{ij} \psi^{(\bar{4}5)}_{kl} \psi^{(2\bar{1}0)}_{ij}
			+\psi^{(\bar{0}1\bar{2}3)}_{kl} \psi^{(\bar{5})}_{ij} \psi^{(5)}_{kl} \psi^{(\bar{3}2\bar{1}0)}_{ij}
			-\psi^{(\bar{0}1\bar{2}3\bar{4})}_{kl} \psi^{(4\bar{3}2\bar{1}0)}_{ij}\right)\; .
		\end{split}	
		\ee
	\end{widetext}
	We use the notation \(\psi^{(a\bar{b}\dots)}_{ij} = \psi^{(a)}_{ij} \psi^{(b)\ast }_{ij}\cdots\) 
	to make the commutator slightly friendlier. 
	
	The operators \({\cal O}_{\pm}(\Pi, \Gamma)\) are built from the components \(\Gamma_{i,j}\) and \(\Pi_{i,j}\). We can 
	define \({\cal O}_{+}(\Pi, \Gamma)\) by locally adding and removing a box in all possible vector positions
	\((j,k)\), whilst \({\cal O}_{-}(\Pi, \Gamma)\)  is defined by locally removing and adding a box in all possible vector positions
	\((j,k)\).
		
	In Appendix~\ref{app:kagome} we build states corresponding to 2- and 3-boxes plane partitions and we also describe 
	the action of the operators \(\Gamma\), \(\Pi\) and \({\cal O}_{\pm}(\Pi, \Gamma)\). Using this construction, one 
	can rewrite the crystal melting Hamiltonian (\ref{eq:Hamiltonian}) as
	\be 
	\label{eq:Hamiltonian2}
	\begin{split}
	H =&  -J \sum_{j,k} \Gamma_{j,k} + \Pi_{j,k}   + \\
	& + V \sum_{j,k} \sqrt{q}\ \Pi_{j,k}  \Gamma_{j,k} +\frac{1}{\sqrt{q}} \Gamma_{j,k}  \Pi_{j,k}\; .
	\end{split}
	\ee
	The ground state of this model has been defined in~\cite{Dijkgraaf:2008ua}. It is the sum of the states corresponding to plane partitions \(\bm{\Lambda}\) weighted by \(q^{|\bm{\Lambda}|}\), where \(|\bm{\Lambda}|=\#\ Boxes (\bm{\Lambda})\). 
	In the notation above we have
	\be  
	\begin{split}
	|\texttt{ground}\rangle & = \sum_{\bm{\Lambda}} q^{\frac{|\bm{\Lambda}|}{2}}|\bm{\Lambda}\rangle \\
	& = \sum_{n} 
	\sum_{ \substack{(j_1, j_2, \dots , j_n) \in \mathbb{Z}^n \\ (k_1, k_2, \dots , j_n) \in \mathbb{Z}^n } }
	 \prod_{a=0}^n q^{\frac{n}{2}} \Gamma_{j_a, k_a} |\bm{\emptyset}\rangle\; ,
	\end{split}
	\ee
	where the first term in the sum, \(n=0\), is simply \(|\bm{\emptyset}\rangle\).
	Therefore, one can verify that \(Z=\langle \texttt{ground} |\texttt{ground}\rangle \) is precisely 
	the MacMahon function~(\ref{eq:part.function}). 
	
	The Hamiltonian~(\ref{eq:Hamiltonian2}) can be thought of as the limit \(J_0=0\) of the more general model
	\be 
	\label{eq:Hamiltonian3}
	H_{hexagon} = J_0\sum_{\langle \vec{r},\vec{r}'\rangle} \psi^\ast(\vec{r})\psi(\vec{r}') + H\; ,
	\ee
	where the first term describes \emph{nearest-neighbor interactions}. In other words, the Hamiltonian~(\ref{eq:Hamiltonian2}) is an interaction Hamiltonian of 6th and 12th degrees, and if we assume a slightly 
	more general setting with lowest-order hopping, we obtain the Hamiltonian~(\ref{eq:Hamiltonian3}).


	\section{Tensor product representation}\label{sec:tensorProd}
	
	In the remainder of this work, we try to understand some universal features of the Hamiltonian (\ref{eq:Hamiltonian}). 
	In other words, we investigate aspects that do not depend on the lattice descriptions above, in particular, 
	its Hilbert space and underlying symmetries.
   The fact that the states in our system are labeled by plane partitions suggests an interpretation in terms of the MacMahon representation of the affine Yangian of \(\widehat{\mathfrak{gl}}(1)\)~\cite{Prochazka:2015deb,Gaberdiel:2017dbk,maulik2012,Tsymbaliuk_2017, Li:2020rij}.
   In this part of the paper we start our program of rewriting and interpreting the Hamiltonian (\ref{eq:Hamiltonian}) in terms of Yangian invariants, which we believe to be the key to understand the integrability properties of the problem.
   We begin by identifying a nontrivial coproduct structure of the operators that we have defined in the previous sections.
	
	Plane partitions can be written as stacks of interlaced integer partitions~\cite{Dijkgraaf:2008ua,Justin2008}, that is
	 \be 
	 \begin{split}
	 & \bm{\Lambda}=\{\Lambda^{(1)}, \Lambda^{(2)}, \dots, \Lambda^{(N)}  | \Lambda^{(a)} 
	 \succ \Lambda^{(a+1)} \}\ , \ \text{where} \\
	 & \Lambda^{(a)}=(\lambda^{(a)}_1, \lambda^{(a)}_2, \dots, \lambda^{(a)}_{\ell_a}  ) 
	 \; , \quad  \lambda^{(a)}_i\geq  \lambda^{(a)}_{i+1} \\
	 & \forall \  i=1,2, \dots , \ell_a\; , \quad a=1,2,\dots, N-1\; .
	 \end{split}
	 \ee
	The number of boxes in the plane partition \(\bm{\Lambda}\) is given by
	 \be 
	 \label{eq:numberboxes01}
	 |\bm{\Lambda}|=\# \text{Boxes} = \sum_{a=1}^N |\Lambda^{(a)}|\; ,
	 \ee
	 in other words, it is the sum of squares of each integer partition layer. In this notation, a generic state \(|\bm{\Lambda}\rangle\) associated to a given plane partition \(\bm{\Lambda}\) can be represented as the tensor product
	 \be 
	 |\bm{\Lambda}\rangle = |\Lambda^{(1)} \rangle\otimes |\Lambda^{(2)} \rangle \otimes \cdots 
	 \otimes |\Lambda^{(N)} \rangle \; ,
	 \ee
	 and the empty plane partition is the product of empty integer partitions
	 \be 
	|\bm{\emptyset}	\rangle = |\emptyset \rangle\otimes |\emptyset \rangle\otimes\cdots \; .
	 \ee
	 As we discuss in Appendix~\ref{app:corres}, the usefulness of this structure is that the integer-partition layers can de described in
	 terms of free fermions with Neveu-Schwarz boundary conditions. \\

	 As we have seen before, the operator \(\Gamma\) adds a box in all consistent 
	 places of a given partition \(\bm{\Lambda}\). Let us denote the space of allowed places as \(\mathcal{Q}^+(\Lambda)\), 
	 therefore
	 \be 
	\Gamma |\bm{\Lambda}\rangle = \sum_{\square \in \mathcal{Q}^+ (\Lambda) } |\bm{\Lambda}+\square\rangle\; .
	 \ee
	 Additionally, the operator \(\Gamma\) can be described in terms of its action on the integer partitions:
	 \bse 
	 \be 
	\begin{split}
		\Gamma |\bm{\Lambda}\rangle & \equiv \Delta(\gamma)\left( \bigotimes_{b=1}^N |\Lambda^{(b)}\rangle \otimes |\emptyset\rangle \right) \\
		 & = \bigoplus_{a=1}^{N+1} \mathrm{R}_a(\gamma)
		 \left( \bigotimes_{b=1}^N |\Lambda^{(b)}\rangle \otimes |\emptyset\rangle \right)\; ,
	\end{split} 
	 \ee
	 where \(\gamma\) is the growth operator on integer partitions. In the definition above, for each layer labeled by \(a\) there exists an action of the integer partition growth operator \(\gamma \in \mathcal{A}\), where \({\cal A}\) is the algebra of operators acting on the space of integer partitions \(H_{(a)}\). In other words, given that \(|\bm{\Lambda}\rangle \in \bigotimes_a H_{(a)}\), the operator \(\Gamma\) can be represented as the coproduct \(\Delta(\gamma)\). Moreover, we assume a decomposition \(\Delta(\gamma) = \bigoplus_{a} \mathrm{R}_a(\gamma)\) that denotes the growth of each integer partition level.
	 
	 A trivial action of the coproduct \(\Delta(\Gamma)\) is given by \(\mathrm{R}_a(\gamma)= \mathbb{1}\otimes \cdots \otimes {\cal R}^{(a)}(\gamma) \otimes \cdots \otimes \mathbb{1}\), where the operator \({\cal R}^{(a)}(\gamma)\) is the (representation of the) growth operator in the tensor product \(a\)-slot. This condition clearly violates the constrained growth of plane partitions; therefore we impose a nontrivial coproduct \(\Delta: {\cal A} \to {\cal A} \otimes {\cal A}\) by
	 \be 
	\mathrm{R}_a(\gamma) := \mathbb{1}\otimes \cdots \otimes {\cal M}^{(a-1)}(\gamma)  \otimes {\cal R}^{(a)}(\gamma) \otimes \cdots \otimes \mathbb{1}\; ,
	 \ee
	  where the operator \({\cal M}^{(a-1)}(\gamma)\) imposes that the \(a\)-slot growth does not violate the plane partition conditions, and its eigenvalues are \(+1\), if the \(a\)-slot integer partition grows consistently, and \(0\) otherwise. The existence of this nontrivial coproduct is the first sign of a quantum group structure underlying the tensor representation above. 	
 	 \ese
	 
	 As an instructive example, consider the configuration 
	 \( \ytableausetup{centertableaux, smalltableaux} | \ytableaushort[*(lightgray)]{2 }\rangle\), with \(\Gamma \ytableausetup{centertableaux, smalltableaux} | \ytableaushort[*(lightgray)]{2 }\rangle = \left|\ytableaushort{2\none}*[*(lightgray)]{2}\right\rangle + \left|\ytableaushort{\none,2}*[*(lightgray)]{1,1}\right\rangle +  | \ytableaushort[*(lightgray)]{3}\rangle  \). In the tensor product notation we have 
 	\bse 
	\be 
	\Gamma \ytableausetup{centertableaux, smalltableaux} | \ytableaushort[*(lightgray)]{2 }\rangle = \bigoplus_{a=1}^3 \mathrm{R}_a(\gamma) \left(
	|\ydiagram[*(lightgray)]{1}\rangle\otimes |\ydiagram[*(lightgray)]{1}\rangle \otimes |\emptyset\rangle\right)\; .
	\ee 	
	The action of \(\mathrm{R}_1(\gamma) \) reads
	\be 
	\begin{split}
	\mathrm{R}_1(\gamma) \left( 
	|\ydiagram[*(lightgray)]{1}\rangle  \otimes |\ydiagram[*(lightgray)]{1}\rangle \otimes |\emptyset\rangle \right) & = \left( \mathcal{R}^{(1)}(\gamma)
	|\ydiagram[*(lightgray)]{1}\rangle \right) \otimes |\ydiagram[*(lightgray)]{1}\rangle \otimes |\emptyset\rangle \\ 
	& = \left( |\ydiagram[*(lightgray)]{2}\rangle + \left|\ydiagram[*(lightgray)]{1,1}\right\rangle \right) \otimes |\ydiagram[*(lightgray)]{1}\rangle \otimes |\emptyset\rangle \\
	& \equiv \left|\ytableaushort{2\none}*[*(lightgray)]{2}\right\rangle + \left|\ytableaushort{\none,2}*[*(lightgray)]{1,1}\right\rangle  
	\end{split}
	\ee
	since the first layer is unconstrained. On the other hand 
	\be 
	\mathrm{R}_2(\gamma) \left( 
	|\ydiagram[*(lightgray)]{1}\rangle  \otimes |\ydiagram[*(lightgray)]{1}\rangle \otimes |\emptyset\rangle \right) = 0\; ,
	\ee
	since any growth of the second layer violates the plane partition conditions. Finally 
	\be 
	\begin{split}
	\mathrm{R}_3(\gamma) \left( 
	|\ydiagram[*(lightgray)]{1}\rangle  \otimes |\ydiagram[*(lightgray)]{1}\rangle \otimes |\emptyset\rangle \right) & = \\
	= |\ydiagram[*(lightgray)]{1}\rangle \otimes &
	\left( \mathcal{M}^{(2)}(\gamma) |\ydiagram[*(lightgray)]{1}\rangle \right) \otimes \left( \mathcal{R}^{(3)}(\gamma) |\emptyset\rangle \right) \\
	= |\ydiagram[*(lightgray)]{1}\rangle \otimes &
	|\ydiagram[*(lightgray)]{1}\rangle \otimes |\ydiagram[*(lightgray)]{1}\rangle \equiv 
	|\ytableaushort[*(lightgray)]{3}\rangle \; .
	\end{split}
	\ee
 	\ese
	 
	 More generally, the first layer can always grow freely as long as it satisfies the integer partition rules. Using the equivalence 
	 between 2D Young diagrams and free fermions, one can represent \({\cal R}^{(1)}(\gamma)\) as
	 \bse
	 \be 
	{\cal R}^{(1)}(\gamma)\equiv \gamma= \sum_{r\in \mathbb{Z}+\frac{1}{2}} \Psi^\ast_{r+1}\Psi_{r}\; .
	 \ee
	Consequently,
	\be 
	\mathrm{R}_1(\gamma) =  \gamma \otimes \mathbb{1}\otimes \cdots \otimes \mathbb{1}\; .
	\ee
	In fact, \(\gamma\) acts as
	 \be 
	 \label{eq:growth}
		\begin{split}
			\gamma & |\Lambda^{(a)}\rangle = 
			| (\lambda_1^{(a)} + 1,\lambda_2^{(a)}, \dots , \lambda_{\ell_a}^{(a)} ) \rangle +\cdots \\
			& + | (\lambda_1^{(a)},\lambda_2^{(a)}, \dots , \lambda_{\ell_a}^{(a)} + 1 ) \rangle + \\
			& + | (\lambda_1^{(a)},\lambda_2^{(a)}, \dots , \lambda_{\ell_a}^{(a)}, 1 ) \rangle\; .
		\end{split} 
	 \ee
	 \ese
	For the subsequent layers, the partitions are subject to the interlacing constraint, then
	 \be 
	 \begin{split}
	 {\cal M}&{}^{(a-1)}  \otimes {\cal R}^{(a)} \left( |\Lambda^{(a-1)}\rangle \otimes |\Lambda^{(a)}\rangle \right) \\ 
	 & =  |\Lambda^{(a-1)}\rangle \otimes  \sum_{ \Lambda^{(a)} +\square \prec \Lambda^{(a-1) }} |\Lambda^{(a)} +\square \rangle  \; , \ a\geq 2\; ,
	\end{split}
	 \ee
	 which means that the resulting partition \(  (\tilde{\Lambda}^{(a)}) \equiv (\Lambda^{(a)} +\square)\) 
	 satisfies \(\tilde{\lambda}^{(a)}_j \leq \lambda^{(a-1)}_j \  \forall\  j=1, \dots, \ell_{a}\). The first, and almost trivial, 
	 observation is that the operators \({\cal M}\) act diagonally, whilst \({\cal R}^{(a)} \) is a constrained growth operator. 
	 
	We propose the form
	 \be 
	 {\cal M}^{(a-1)}\otimes {\cal R}^{(a)} = \sum_i  {\cal M}_i^{(a-1)}\otimes {\cal R}_i^{(a)}\; ,
	 \ee
	 where \({\cal R}_i^{(a)}\) acts as
	 \be 
	 \begin{split}
	 {\cal R}_i^{(a)}| (\lambda_1^{(a)}, \dots , & \lambda_i^{(a)},  \dots) \rangle = \\
	 | (\lambda_1^{(a)}, & \dots , \lambda_i^{(a)} + 1,  \dots) \rangle.
	\end{split}
	\ee 
	 The operator \({\cal M}^{( a-1)}_i \) tests whether the partition 
	 \((\Lambda^{(a)} +\square)\) interlaces \((\Lambda^{(a-1)})\), and it can be defined as
	  \bse
	  \be
	  \begin{split}
	  	{\cal M}^{(a-1)}_i | (\lambda_1^{(a-1)}, & \dots , \lambda_i^{(a-1)},  \dots) \rangle = \\
	  	& r_{i, a} | (\lambda_1^{(a-1)}, \dots , \lambda_i^{(a-1)}, \dots ) \rangle\; ,
	  \end{split}
	  \ee
	  where \(i=1,\dots, \ell_a, \ell_a+1\) and 
	  \be 
		r_{i, a} = \left\{ 
		\begin{array}{lll}
		0 && \text{if } \ \ 	\lambda_i^a+1> \lambda_i^{a-1}\\
		1 && \text{otherwise} 
		\end{array}
		\right.  \; .
	  \ee
  	  \ese
	  Observe that the operator 
	  \({\cal M}^{( a-1)} \) is nonlocal, since it acts on the \((a-1)\)-slot but additional information on the states of the 
	  \(a\)-slot is required. It would very be interesting to find representation for the operators \( {\cal M}^{(a-1)}\otimes {\cal R}^{(a)}\), for \(a\geq 2\), using the free fermions formalism.

	\medskip
	Similarly, the operator \(\Pi\) removes a box in all consistent places, denoted by \(\mathcal{Q}^-(\Lambda)\), in a given 
	partition \(\bm{\Lambda}\). Then
	\be 
	\Pi |\bm{\Lambda}\rangle = \sum_{\square \in \mathcal{Q}^- (\Lambda) } |\bm{\Lambda}-\square\rangle\; .
	\ee
	As before, we write 
	\bse 
	\be 
	\begin{split}
		\Pi |\bm{\Lambda}\rangle & \equiv \Delta(\pi)\left( \bigotimes_{b=1}^N 
		|\Lambda^{(b)}\rangle \otimes |\emptyset\rangle \right) \\
		& = \bigoplus_{a=1}^N \mathrm{S}_a(\pi)
		\left( \bigotimes_{b=1}^N |\Lambda^{(b)}\rangle \otimes |\emptyset\rangle \right)\; ,
	\end{split} 
	\ee
	where \(\pi\) acts on integer partitions. In addition
	\be 
	\mathrm{S}_a(\pi) = \mathbb{1}\otimes \cdots \otimes {\cal S}^{(a)}(\pi) \otimes {\cal N}^{(a+1)}(\pi)\otimes \mathbb{1} \cdots \otimes \mathbb{1}\; .
	\ee
	\ese
	
	The final layer can shrink freely as long as it satisfies the integer partition rules, therefore
	\be 
	{\cal S}^{(N)}(\pi)\equiv \pi= \sum_{r\in \mathbb{Z}+\frac{1}{2}} \Psi_{r}^\ast \Psi_{r+1}
	\ee
	and 
	\be 
	\label{eq:shrink}
	\begin{split}
		\pi & |\Lambda^{(a)}\rangle = 
		| (\lambda_1^{(a)} - 1,\lambda_2^{(a)}, \dots , \lambda_{\ell_a}^{(a)} ) \rangle +\cdots \\
		& + | (\lambda_1^{(a)},\lambda_2^{(a)}, \dots , \lambda_{\ell_a}^{(a)} - 1 ) \rangle .
	\end{split} 
	\ee
	For the precedent layers, the partitions are subject to the interlacing constraint, therefore
	\be 
	\begin{split}
	{\cal S}^{(a)} \otimes & {\cal N}^{(a+1)} \left( |\Lambda^{(a)}\rangle \otimes |\Lambda^{(a+1)}\rangle\right) \\ 
	& = \sum_{ \Lambda^{(a)} -\square \succ \Lambda^{(a+1) }} |\Lambda^{(a)} -\square \rangle 
	\otimes |\Lambda^{(a+1)}\rangle \; , \ a < N\; .
	\end{split}
	\ee	
	We propose the form 
	\be 
	{\cal S}^{(a)} (\pi) \otimes {\cal N}^{(a+1)} = \sum_i {\cal S}^{(a)}_i (\pi) \otimes {\cal N}^{(a+1)}_i\; ,
	\ee
	where \({\cal N}^{(a+1)} \) is defined by 
	\bse
	\be
	\begin{split}
		{\cal N}^{(a+1)}_i | (\lambda_1^{(a+1)}, & \dots , \lambda_i^{(a+1)},  \dots   ) \rangle = \\
		& s_{i, a} | (\lambda_1^{(a+1)}, \dots , \lambda_i^{(a+1)}, \dots ) \rangle\; ,
	\end{split}
	\ee
	where \(i=1,\dots, \ell_a, \ell_a\) and 
	\be 
	s_{i, a} = \left\{ 
	\begin{array}{lll}
		0 && \text{if } \ \ 	\lambda_i^a-1< \lambda_i^{a+1}\\
		1 && \text{otherwise} 
	\end{array}
	\right.  \; .
	\ee
	\ese
	It would be interesting to study the fermionic representation of the operators \({\cal M}\) and 
	\({\cal N}\) and the quantum group structure of the operators \(\Gamma\) and \(\Pi\) in this context. On the other hand,
	in the next section we will see that the growth of plane partitions is better described in terms of the Yangian generators, 
	and in this new context some of the underlying algebraic structure becomes more transparent. It is worth 
	mentioning that although the Yangian algebra has its defining coproduct structure, it is different from the coproduct structure we consider in this section.


	\section{ Fermion-Boson Duality \& Affine Yangian}
	\label{sec:fb-corr}
	
	The important observation of the previous section is the nontrivial coproduct structure of the operators 
	\(\Gamma\) and \(\Pi\) in the tensor product representation. In this section, we study other algebraic structure 
	of these operators in terms of the affine Yangians of \(\mathfrak{gl}(1)\), and we also define a fermion-boson 
	correspondence for the plane partition states. 
	
	In Appendix~\ref{app:corres} we review the necessary 
	tools we use in this section. In order to define a fermion boson-correspondence for plane partition states, let us write the states
	\(|\bm{\Lambda}\rangle\), that we call \emph{fermionic plane partition state}, as
	\bse
	\be 
	\begin{split}
		|\bm{\Lambda}\rangle & = \bigotimes_{a=1}^N \sum_{\vec{k}^{(a)}}	
		\frac{\chi_{\Lambda^{(a)}}[C(\vec{k}^{(a)})]}{z_{\vec{k}^{(a)}}}
		|\vec{k}^{(a)}\rangle\\
		& = \sum_{\vec{K}}\frac{\chi_\Lambda[C(\vec{K})] }{Z_{\vec{K}}} |\vec{K}\rangle,
	\end{split}
	\ee
	where we use 
	\be 
	|\vec{K}\rangle = |\vec{k}^{(1)}\rangle \otimes \cdots \otimes |\vec{k}^{(N)}\rangle ,
	\ee 
	and 
	\be 
	\label{eq:characters}
	\chi_\Lambda[C(\vec{K})] = \chi_{\Lambda^{(1)}}[C(\vec{k}^{(1)})]\cdots 
	\chi_{\Lambda^{(N)}}[C(\vec{k}^{(N)})]\; 
	\ee
	\ese
	is a product of characters of the symmetric group. Using the results of Appendix~\ref{app:corres} and 
	equation~(\ref{eq:numberboxes01}), the number of boxes in \(\bm{\Lambda}\) is simply
	\be 
	\label{eq:numberboxes02}
	|\bm{\Lambda}|	= \sum_{a=1}^N \sum_j j k_j^{(a)}\; .
	\ee

	The states \(|\vec{k}\rangle\) can be labeled in terms of partitions themselves, that is
	\be 
	|\vec{k}\rangle = |\dots ,\underbrace{2,\dots, 2}_{k_2 \ \text{times}}, \underbrace{1,\dots, 1}_{k_1 \ \text{times}}\rangle \; ,
	\ee
	and we denote these states as \(|\vec{k}\rangle \equiv |\mathbb{N}_{\vec{k}}\rrangle\). For example, the state associated to the 
	vector \(\vec{k}=(0,1,2,1,0,0,\dots)\) is \(\alpha_{-4}\alpha_{-3}^2\alpha_{-2}|0\rangle\) and we denote it as a \emph{bosonic 
	plane partition state} \(|(4,3,3,2)\rrangle\). Using the Frobenius formula (\ref{eq:frobenius}), we can write the plane partition state 
	in terms of \(|\mathbb{N}_{\vec{k}^{(a)}}\rrangle\),
	\be 
	\begin{split}
		\label{eq:btransf01}
		|\bm{\Lambda}\rangle & = \bigotimes_{a=1}^N \sum_{\vec{k}^{(a)}}	
		\frac{\chi_{\Lambda^{(a)}}[C(\vec{k}^{(a)})]}{z_{\vec{k}^{(a)}}}
		|\mathbb{N}_{\vec{k}^{(a)}}\rrangle\\
		& = \sum_{\vec{K}}\frac{\chi_\Lambda[C(\vec{K})] }{Z_{\vec{K}}} |\bm{\Lambda}(\mathbb{N}_{\vec{K}})\rrangle\; .
	\end{split}
	\ee
	Its inverse is
	\be 
	\label{eq:btransf02}
	\begin{split}
		|\bm{\Lambda}(\mathbb{N}_{\vec{K}})\rrangle = \sum_{\Lambda} \chi_\Lambda[C(\vec{K})] |\bm{\Lambda}\rangle\; .
	\end{split}
	\ee
	The relations (\ref{eq:btransf01}) and (\ref{eq:btransf02}) define the plane partition fermion-boson correspondence. 
	From (\ref{eq:numberboxes02}), it is clear that \(|\bm{\Lambda}|=|\bm{\Lambda}(\mathbb{N}_{\vec{K}}) |\), therefore, the 
	correspondence between the plane partition basis preserves the \emph{number of boxes grading}. It can be seen 
	form the fact that the only nontrivial characters associated to the Young diagram \(\bm{\Lambda}\) are those related to the 
	conjugacy classes of \(|\bm{\Lambda}|\). In Appendix~\ref{app:basis} we write these relations for a few states. 
	
	We would like to represent the growth operators, and consequently the Hamiltonian (\ref{eq:Hamiltonian}), 
	in terms of the new bosonic basis \(|\bm{\Lambda}(\mathbb{N}_{\vec{k}^{(a)}})\rrangle\). It is easy to see that since the map preserves the number of boxes
	\bse
	\be 
	\label{eq:creation}
	\begin{split}
	&\Gamma |\bm{\Lambda}(\mathbb{N}_{\vec{K}})\rrangle = \sum_{\Lambda}\chi_{\Lambda}[C(\vec{K})] \sum_{\square \in {\cal Q}^+(\Lambda), \vec{L}} \frac{\chi_{\Lambda +\square}[C(\vec{L})]}{Z_{\vec{L}}} 
	|\bm{\Lambda}(\mathbb{N}_{\vec{L}})\rrangle\\
	& \equiv \sum_{\square \in \mathcal{Q}^+(\Lambda(\mathbb{N}_{\vec{K}})) }
	{\cal A}_+\left( \bm{\Lambda}(\mathbb{N}_{\vec{K}})\mapsto \bm{\Lambda}(\mathbb{N}_{\vec{K}}) + \square  \right) | \bm{\Lambda}(\mathbb{N}_{\vec{K}}) + \square \rrangle\; ,
	\end{split}
	\ee
	where \(|\bm{\Lambda}(\mathbb{N}_{\vec{L}})|= |\bm{\Lambda}(\mathbb{N}_{\vec{K}})|+1\) and \(\sum_{\vec{L}}=\sum_{\square \in \mathcal{Q}^+(\Lambda(\mathbb{N}_{\vec{K}})) }\). Therefore, we have the amplitudes
	\be 
	\label{eq:ampcreat}
	\begin{split}
		{\cal A}_+ & \left( \bm{\Lambda}(\mathbb{N}_{\vec{K}})\mapsto \bm{\Lambda}(\mathbb{N}_{\vec{K}}) + \square  \right) = \\
		& =\sum_{\Lambda} \sum_{\square \in {\cal Q}^+(\Lambda) } 
		\chi_{\Lambda}[C(\vec{K})]  \frac{\chi_{\Lambda +\square}[C(\vec{L})]}{Z_{\vec{L}}},
	\end{split}
	\ee
	where \(\vec{L}\) is the conjugacy class associated to the plane partition 
	\(\bm{\Lambda}(\mathbb{N}_{\vec{L}}) = \bm{\Lambda}(\mathbb{N}_{\vec{K}}) + \square\).
	\ese	 
		
	Similarly, 
	\bse
	\be 
	\label{eq:annihilation}
	\begin{split}
		&\Pi |\bm{\Lambda}(\mathbb{N}_{\vec{K}})\rrangle = \sum_{\Lambda}\chi_{\Lambda}[C(\vec{K})] \sum_{\square \in {\cal Q}^-(\Lambda), \vec{J}} \frac{\chi_{\Lambda -\square}[C(\vec{J})]}{Z_{\vec{J}}} 
		|\bm{\Lambda}(\mathbb{N}_{\vec{J}})\rrangle\\
		& \equiv \sum_{\square \in \mathcal{Q}^-(\Lambda(\mathbb{N}_{\vec{K}})) }
		{\cal A}_-\left( \bm{\Lambda}(\mathbb{N}_{\vec{K}})\mapsto \bm{\Lambda}(\mathbb{N}_{\vec{K}}) - \square  \right) | \bm{\Lambda}(\mathbb{N}_{\vec{K}}) - \square \rrangle\; ,
	\end{split}
	\ee
	where \(|\bm{\Lambda}(\mathbb{N}_{\vec{J}})|= |\bm{\Lambda}(\mathbb{N}_{\vec{K}})|-1\) and \(\sum_{\vec{J}}=\sum_{\square \in \mathcal{Q}^-(\Lambda(\mathbb{N}_{\vec{K}})) }\). The amplitudes are
	\be 
	\label{eq:ampannih}
	\begin{split}
		{\cal A}_- & \left( \bm{\Lambda}(\mathbb{N}_{\vec{K}})\mapsto \bm{\Lambda}(\mathbb{N}_{\vec{K}}) - \square  \right) = \\
		& =\sum_{\Lambda} \sum_{\square \in {\cal Q}^-(\Lambda) } 
		\chi_{\Lambda}[C(\vec{K})]  \frac{\chi_{\Lambda -\square}[C(\vec{J})]}{Z_{\vec{J}}},
	\end{split}
	\ee
	where \(\vec{J}\) is the conjugacy class associated to the plane partition 
	\(\bm{\Lambda}(\mathbb{N}_{\vec{J}}) = \bm{\Lambda}(\mathbb{N}_{\vec{K}}) - \square\).
	\ese	 
	We give the amplitudes for the 0-, 1- and 2-box transitions in Appendix~\ref{app:amplitudes}.
	
	Working with the (bosonic) plane partition states \(|\mathbb{N}_{\vec{K}}\rrangle\) seems to be much more involved, 
	but it, in fact, has some conceptual advantages. In particular, the action of the growth operators 
	on \(|\mathbb{N}_{\vec{K}}\rrangle\) suggests a relation between these states and the MacMahon 
	representation of the affine Yangian of \(\widehat{\mathfrak{gl}}(1)\)~\cite{Prochazka:2015deb,Gaberdiel:2017dbk,maulik2012,Tsymbaliuk_2017, Li:2020rij}. 
	More specifically, we have the following conjecture:
	
	\begin{framed}
	\emph{The growth operators \(\Gamma\), \(\Pi\) and \({\cal O}_\pm (\Gamma, \Pi)\) are 
	linear combinations of the generators of the algebra \({\cal Y}[\widehat{\mathfrak{gl}}(1)]\),
	and the states \(|\mathbb{N}_{\vec{K}}\rrangle\) fulfill the \({\cal Y}[\widehat{\mathfrak{gl}}(1)]\) 
	MacMahon representation.}
	\end{framed}
	
	The algebra \({\cal Y}[\widehat{\mathfrak{gl}}(1)]\) is an associative algebra defined by the 
	generators \(\{e_j, f_j, \varphi_j\ | \ j\geq 0\}\) and a set of (anti-) commutations that can be found in~\cite{Prochazka:2015deb}.
	One of the most important features of the affine Yangian of \(\widehat{\mathfrak{gl}}(1)\) is its triangular decomposition 
	\( {\cal Y}[\widehat{\mathfrak{gl}}(1)] =  {\cal Y}^+\oplus {\cal B}\oplus {\cal Y}^-\) which is generated, respectively, by \(e_j\), \(\varphi_j\) and \(f_j\). 
	
	As a first evidence in favor of our conjecture, it has been shown that in the MacMahon representation studied by Proch\'{a}zka~\cite{Prochazka:2015deb}, the operators \(e_j\) and \(f_j\) act precisely as in 
	(\ref{eq:creation}) - (\ref{eq:annihilation}), that is
	\be 
	\begin{split}
		e_j | \bm{\Lambda}\rrangle & = \sum_{\square\in {\cal Q}^+}E_j(\bm{\Lambda} \mapsto \bm{\Lambda} + \square) |\bm{\Lambda}+\square\rrangle\; ,\\
		f_j | \bm{\Lambda}\rrangle & = \sum_{\square\in {\cal Q}^-}F_j(\bm{\Lambda} \mapsto \bm{\Lambda} - \square) |\bm{\Lambda}-\square\rrangle .
	\end{split}
	\ee
	Furthermore, the character of this representation is precisely the MacMahon function (\ref{eq:part.function}), and 
	the operators \(\varphi_j\) act diagonally as
	\be 
	\varphi_j |\bm{\Lambda}\rrangle = \varphi_{j, \Lambda}  |\bm{\Lambda} \rrangle\; .
	\ee
	Finally, if the conjectured correspondence between the generators of \({\cal Y}[\widehat{\mathfrak{gl}}(1)] \) and 
	with the growth operators \(\Gamma\) and \(\Pi\) is indeed correct, we immediately conclude 
	that \({\cal O}_\pm(\Pi , \Gamma) \in {\cal B} \) since they can be written as linear combinations of \(\varphi_j\). 
	In other words, both systems are precisely the 
	same once we identify the amplitudes \({\cal A}_\pm\) with linear combinations of the 
	coefficients \(E_j\) and \(F_j\). We then conclude that 
	the Hamiltonian (\ref{eq:Hamiltonian}) can be written as
	\be 
	\begin{split}
		\label{eq:Hamiltonian4}
		&H = - J\sum_{j \in \mathbb{Z}_+} a_j \left( e_j - f_j\right) \\
		&+ V\sum_{j \in \mathbb{Z}_+} \left( \sqrt{q} \ b_j+ \frac{1}{\sqrt{q}}\ c_j  \right)\varphi_j + V(c_0 \varphi_0 +c_1 \varphi_1 )\; ,
	\end{split}
	\ee
	where the coefficients \(a_j\), \(b_j\) and \(c_j\) are necessary to define the equivalence, including combinatorial terms, without introducing new constraints to the Yangian \({\cal Y}[\widehat{\mathfrak{gl}}(1)]\). Observe that we write the generators \(\varphi_0\) and \(\varphi_1\) separately since they are central elements.


	\section{Conclusions and Outlook}\label{sec:conclusions}
	
	In this work we have studied the 3D quantum crystal melting of~\cite{Dijkgraaf:2008ua}, governed by the Hamiltonian (\ref{eq:Hamiltonian}) in various guises.
	An unanswered question is whether or not this system is quantum integrable, as is the case for its 2D cousin. In the current paper, we push this study further. We have rewritten the dimer model  
	as a fermionic system with 6th-order interactions in a Kagome lattice. We can think of this relation as a duality, where 
	the extended object (the dimer) in the hexagonal lattice corresponds to a particle in the dual Kagome lattice. This description and its possible advantages have not yet been fully explored. 
	
	Using the tensor product representation of the plane partitions, we have argued that the growth operators have a
	nontrivial coproduct, which may point to the existence of underlying quantum group structures. In fact, it is interesting to explore
	the conjectured relation between this structure and the coproduct of the affine Yangian of \(\widehat{\mathfrak{gl}}(1)\). 
	Additionally, using the fermion-boson correspondence, we have been able to extend the duality to the 3D plane 
	partition states. Putting all these facts together, we managed to write the crystal melting Hamiltonian in terms of Yangian generators, although the conjecture is yet to be proved. 
	\medskip
	
	There is a myriad of interesting problems still to be addressed. As mentioned, the Kagome 
	lattice description is still poorly understood and it may be an interesting description from the viewpoint of condensed matter physics. 
	It would be very desirable to study the dynamics of the more general Hamiltonian (\ref{eq:Hamiltonian3}), 
	and see how the states classified by plane partitions emerge from the limit \(J_0=0\). 
	 
	We have not specified details of the Yangian \({\cal Y}[\widehat{\mathfrak{gl}}(1)]\), but 
	we should mention that we have been able to use its representation because there are isomorphisms relating
	the Yangian \({\cal Y}[\widehat{\mathfrak{gl}}(1)]\), the \({\cal W}_{1+\infty}\)
	and the spherical Hecke algebra SH\(^c\), see~\cite{Prochazka:2015deb,Gaberdiel:2017dbk}. 
	More specifically, the linear \({\cal W}_{1+\infty}\)
	has a well-known representation in terms of free fermions, and in Section~(\ref{sec:fb-corr}) we have used the 
	Heisenberg subalgebra \(\hat{u}(1)\subset {\cal W}_{1+\infty}\) to define the bosonic plane partition states. 
	
	We should also observe that the maps relating the Heisenberg and the Yangian generators are of the form
	\(\alpha_{-m}\propto \text{ad}_{e_1}^{m-1}e_0\) and \(\alpha_{m}\propto \text{ad}_{f_1}^{m-1}f_0\); and that the 
	Heisenberg subalgebra is the only ingredient we used in the definitions of the bosonized expressions. 
	Therefore, it is possible that the only nontrivial coefficients in 
	(\ref{eq:Hamiltonian4}) are \(a_0\) and \(a_1\). 
	Finally, we have specifically realized the construction in terms of the linear \({\cal W}_{1+\infty}\), but one can 
	try to generalize it to the nonlinear \({\cal W}_{1+\infty}\) domain assuming, for example, that 
	the Hamiltonian (\ref{eq:Hamiltonian4}) is the defining problem.
	
	In conclusion, despite the evidence in favor of the integrability of the Hamiltonian (\ref{eq:Hamiltonian}), we have not yet been able to close this question. Let us hope nature does not disappoint us.

	\section{Acknowledgments}
	
	T.A. is supported by the Swiss National Science Foundation under Grant No. \textsc{PP00P2\_183718/1}. 
	D.O. acknowledges partial support by the NCCR 51nf40–141869 “The Mathematics of
	Physics” (Swissmap). The work of S.R. is supported by the Swiss National Science Foundation under Grant No. \textsc{PP00P2\_183718/1}.


	\appendix
	
	\section{2D Fermion-Boson Duality}
	\label{app:corres}
	
	Bosonization in two dimensions can be rephrased as a duality between the fermionic and bosonic Hilbert spaces.
	In this section we would like to review the necessary ingredients for the construction of the fermion-boson duality 
	for plane partitions of Section~\ref{sec:fb-corr}, 
	in this section we follow~\cite{Jimbo:1983if,Ketov:1995yd,Babelon2003,Okounkov:2003sp,Dijkgraaf:2008ua,Marino:2005sj}.

	Neveu-Schwarz free fermions states can be classified in terms of integer partitions. 
	Using the algebra
	\be 
	\begin{split}
		& \{\psi_m, \psi_n\}= \{\psi^\ast_m, \psi^\ast_n\} = 0, \\
		& \{\psi^\ast_m, \psi_n\} = \delta_{mn} \; ,
	\end{split}
	\ee 
	the fermionic state \(|\lambda\rangle\) in the Hilbert space \({\cal H}\) is given by
	\be 
	|\lambda\rangle = \varepsilon \prod_{i=1}^{r} \psi_{-m_i-1/2}\psi_{-n_i-1/2}^\ast |0\rangle\; ,
	\ee
	where the vacuum \(|0\rangle\) is defined by the conditions 
	\be 
	\psi_n |0\rangle =\psi^\ast_n |0\rangle =0 \; \forall \ n>0\; ,
	\ee
	and \( \varepsilon=(-1)^{\sum^r n_i +r(r-1)/2 }\).
	
	One can now associate the plane partition 
	\(\lambda=(\lambda_1, \lambda_2, \dots)\) to the state above, which in Frobenius coordinates reads
	\be 
	\lambda=\{(m_i| n_i)| i=1, \dots , r\}\; .
	\ee
	The number of boxes associated to this Young diagram is
	\be 
	|\lambda|=\sum_j \lambda_j =\sum_i (m_i + n_i)\; .
	\ee
	
	In two dimensions, one can expand the same Hilbert space \({\cal H}\) in terms of a chiral bosonic field \( \partial \phi(z)\) defined by the components
	\bse 
	\be 
	\alpha_m = \sum_{j\in \mathbb{Z}+1/2}\normord \psi_{-j}\psi^\ast_{j+m} \normord \; ,
	\ee
	which satisfy the Heisenberg algebra
	\be 
	[\alpha_m, \alpha_n]=m \delta_{m+n,0}\; .
	\ee
	\ese
	Using these operators, states in \({\cal H}\) are of the form 
	\be 
	|\vec{k}\rangle \equiv |k_1, k_2, \dots \rangle = \prod_{j=1}^{\infty} (\alpha_{-j})^{k_j}|0\rangle\; .
	\ee
	
	The relation between the two bases \(|\vec{k}\rangle\) and \(|\lambda\rangle\) is given by 
	\bse
	\be 
	\label{eq:basis}
	|\lambda\rangle = \sum_{\vec{k}} \frac{\chi_\lambda[C(\vec{k})]}{z_{\vec{k}}} |\vec{k}\rangle\; ,
	\ee
	and its inverse is 
	\be 
	|\vec{k}\rangle = \sum_{\lambda} \chi_\lambda[C(\vec{k})] |\lambda\rangle \; ,
	\ee
	\ese
	where \(z_{\vec{k}} = \prod_{j=1}^\infty k_j! j^{k_j}\), and \(\chi_\lambda[C(\vec{k})]\) is the character 
	of the symmetric group \(\mathfrak{S}_n\) evaluated in the representation \(\lambda\) and 
	conjugacy class \(C(\vec{k})\). Moreover, the number of boxes in the partition \(\lambda\) is the level 
	of the state \(|\vec{k}\rangle\), that is 
	\be 
	|\lambda|= \sum_{j\geq 1} j k_j\; . 
	\ee
	
	The characters can be easily calculated using the 
	\emph{Frobenius character formula}~\cite{Fulton}
	\bse
	\be 
	\label{eq:frobenius}
	\chi_{\lambda}[C(\vec{k})] = \left[ 
	\Delta(\vec{x}) \prod_{j=1}^{r} P_j(\vec{x})^{k_j} 
	\right]_{(\ell_1, \dots, \ell_m)}\; ,
	\ee
	with
	\be 
	\Delta(\vec{x}) = \prod_{i<j}(x_i-x_j)\; , \quad P_j(\vec{x})=\sum_{i=1}^mx_i^j\; .
	\ee
	\ese
	The vector \(\vec{k}\) is assumed to have a finite number of nonzero terms, that is
	\(\vec{k} = (k_1, \dots, k_r)\). The auxiliary coordinates are
	\(\vec{x} = (x_1, \dots, x_m)\) and \(m\) is the number rows of the Young diagram 
	\(\lambda = (\lambda_1, \dots , \lambda_m)\). Furthermore, we also have the coefficients 
	\be 
	\ell_j = \lambda_j + m -j\; ,\ \ j=1,\dots, m\; ,
	\ee
	and given a generic polynomial \(f(\vec{x})\), we denote
	\be 
	\left. f(\vec{x})\right|_{(\ell_1, \dots, \ell_m)}:=\text{coeff. of} \ \ x_1^{\ell_1}\cdots x_m^{\ell_m}\; .
	\ee
	
	For completeness, we give here the character tables we use in the text:
	\be 
	\label{eq:char}
	\chi_{(1)}[C(1,0,\dots)]=1\; ,
	\ee
	
	\begin{center}
		\begin{tabular}{ crr } 
			\toprule
			& \((0,1,0,\dots)\) & \((2,0,\dots)\) \\
			\midrule
			\(\chi_{(1,1)}\) & \(-1\) & \(1\) \\ 
			\(\chi_{(2)}\) & \(1\) & \(1\) \\ 
			\bottomrule
		\end{tabular}
	\end{center}
	
	\begin{center}
		\begin{tabular}{ crrr } 
			\toprule
			& \((0,0,1,\dots)\) &  \((1,1,0,\dots)\) & \((3,0,0,\dots)\) \\
			\midrule
			\(\chi_{(1,1,1)}\) & \(1\) & \(-1\) & \(1\) \\ 
			\(\chi_{(2,1)}\) & \(-1\) & \(0\) & \(2\) \\ 
			\(\chi_{(3)}\) & \(1\) & \(1\) & \(1\) \\ 
			\bottomrule
		\end{tabular}
	\end{center}


	\section{Kagome lattice Hamiltonian}
	\label{app:kagome}
	
	In this section we use the operators \(\Gamma\) and \(\Pi\) to argue that the Hamiltonian (\ref{eq:Hamiltonian2}) 
	in the Kagome lattice is equivalent to (\ref{eq:Hamiltonian}).
	
	Let us first write some states. The 2-boxes configurations are
	\bse
	\be 
	\ytableausetup{centertableaux, smalltableaux}
	\left| \ydiagram[*(lightgray)]{1,1}\right\rangle = \Gamma_{1,0} | \ydiagram[*(lightgray)]{1}\rangle, \
	\ytableausetup{centertableaux, smalltableaux}
	\left| \ydiagram[*(lightgray)]{2}\right\rangle = 
	\Gamma_{0,1} | \ydiagram[*(lightgray)]{1}\rangle, \
	\ytableausetup{centertableaux, smalltableaux}
	| \ytableaushort[*(lightgray)]{2 }\rangle = \Gamma_{-1,-1} | \ydiagram[*(lightgray)]{1}\rangle\; ,
	\ee
	and inversely,
	\be 
	\ytableausetup{centertableaux, smalltableaux}
	| \ydiagram[*(lightgray)]{1}\rangle = \Pi_{1,0} \left| \ydiagram[*(lightgray)]{1,1}\right\rangle , \
	| \ydiagram[*(lightgray)]{1}\rangle = \Pi_{0,1}\ytableausetup{centertableaux, smalltableaux}
	\left| \ydiagram[*(lightgray)]{2}\right\rangle , \
	| \ydiagram[*(lightgray)]{1}\rangle = \Pi_{-1,-1}\ytableausetup{centertableaux, smalltableaux}
	| \ytableaushort[*(lightgray)]{2 }\rangle\; .
	\ee
	\ese
	
	Additionally, the 3-box configurations are
	\bse
	\be 
	\nonumber
	\ytableausetup{centertableaux, smalltableaux} \left| \ydiagram[*(lightgray)]{1,1,1}\right\rangle = \Gamma_{2,0} \left| \ydiagram[*(lightgray)]{1,1}\right\rangle\; , \quad 
	\left| \ydiagram[*(lightgray)]{3}\right\rangle = \Gamma_{0,2} \left| \ydiagram[*(lightgray)]{2}\right\rangle \; ,
	\ee
	\be
	\nonumber
	\left| \ydiagram[*(lightgray)]{1,2}\right\rangle = 
	\Gamma_{0,1} \left| \ydiagram[*(lightgray)]{1,1}\right\rangle = 
	\Gamma_{1,0} \left| \ydiagram[*(lightgray)]{2}\right\rangle\; , 
	\ee
	\be 
	| \ytableaushort[*(lightgray)]{3}\rangle = \Gamma_{-2,-2} | \ytableaushort[*(lightgray)]{2}\rangle \; ,
	\ee
	\be
	\nonumber
	\left|\ytableaushort{\none,2}*[*(lightgray)]{1,1}\right\rangle = \Gamma_{-1,-1} \left| \ydiagram[*(lightgray)]{1,1}\right\rangle  \; , 
	\quad
	\left|\ytableaushort{2\none}*[*(lightgray)]{2}\right\rangle = \Gamma_{-1,-1} \left| \ydiagram[*(lightgray)]{2}\right\rangle
	\ee
	and inversely,
	\be 
	\nonumber
	\left| \ydiagram[*(lightgray)]{1,1}\right\rangle =
	\Pi_{2,0} \ytableausetup{centertableaux, smalltableaux} \left| \ydiagram[*(lightgray)]{1,1,1}\right\rangle
	= \Pi_{0,1}  \left| \ydiagram[*(lightgray)]{1,2}\right\rangle = 
	\Pi_{-1,-1}\left|\ytableaushort{\none,2}*[*(lightgray)]{1,1}\right\rangle \; ,
	\ee
	\be
	\nonumber
	\left| \ydiagram[*(lightgray)]{2}\right\rangle = \Pi_{0,2}  \left| \ydiagram[*(lightgray)]{3}\right\rangle =
	\Pi_{1,0} \left| \ydiagram[*(lightgray)]{1,2}\right\rangle =
	\Pi_{-1,-1}\left|\ytableaushort{2\none}*[*(lightgray)]{2}\right\rangle\; , 
	\ee
	\be 
	| \ytableaushort[*(lightgray)]{2}\rangle = \Pi_{-2,-2} | \ytableaushort[*(lightgray)]{3}\rangle= 
	\Pi_{0,1}\left|\ytableaushort{2\none}*[*(lightgray)]{2}\right\rangle = 
	\Pi_{1,0}\left|\ytableaushort{\none,2}*[*(lightgray)]{1,1}\right\rangle  \; .
	\ee
	\ese
	
\begin{figure}[h]
		\centering	
		\includegraphics[width=2.5cm]{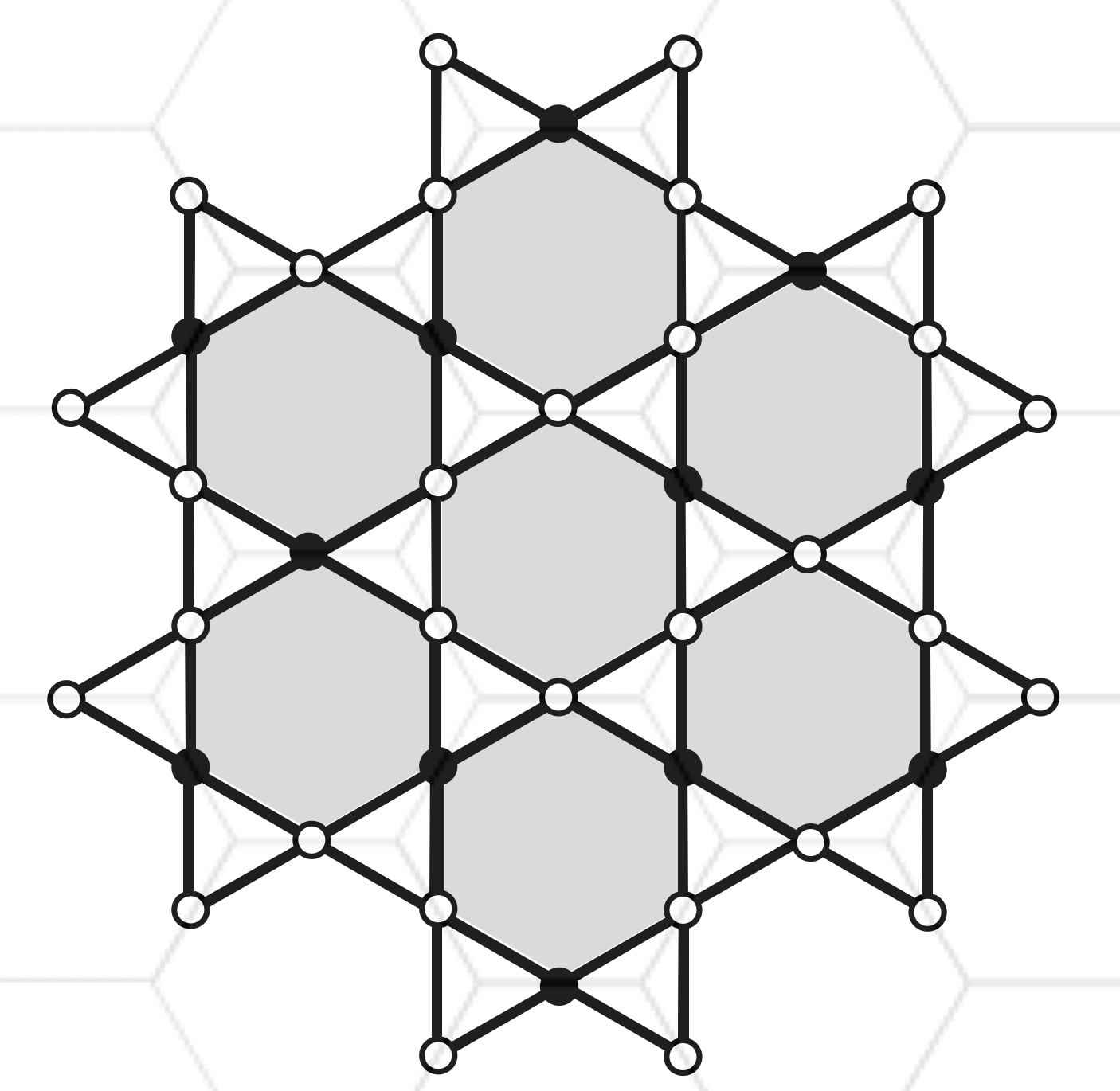}
		\caption{A 5-box configuration}
		\label{fig:5-boxes}
	\end{figure}

	Terms with more boxes can be written in a similar manner. In figure~\ref{fig:5-boxes} we give the 5-box configuration 
	\be 
	\ytableausetup{centertableaux, smalltableaux}\left| \ytableaushort{\none\none,2\none}*[*(lightgray)]{2,2}\right\rangle
	=\Gamma_{-1,-1}\Gamma_{1,1}\Gamma_{1,0}\Gamma_{0,1}\Gamma_{0,0}|\emptyset\rangle 
	\ee 
	in the dual lattice.
	
	In summary, we write the kinetic term of the Hamiltonian (\ref{eq:Hamiltonian}) as
	\be 
	\sum |\blo\rangle \langle \spp |  + \sum |\spp\rangle \langle \blo | = 
	\sum_{j,k} \Gamma_{j,k} + \Pi_{j,k}\; .
	\ee
	
	For the potential term, we need to write operators that give the number of places where we can add a box, and the number of boxes 
	that can be consistently removed. These operators can be written as products of \(\Gamma\) and \(\Pi\). 
	For example, given the empty configuration, it is easy to see that
	\be 
	\sum_{i,j;k,l} \Gamma_{i,j} \Pi_{k,l}  |\emptyset\rangle = 0\; , \quad 
	\sum_{i,j;k,l} \Pi_{i,j}  \Gamma_{k,l}|\emptyset\rangle = |\emptyset\rangle\; ,
	\ee
	that, naturally, means that we have one available place to add a box, and we do not have any box to be removed. 
	
	For two boxes, we see that there must be relations between the two sets of indices \((i,j)\) and \((k,l)\). Initially we have 
	\be 
	\sum_{i,j} \Gamma_{0,0} \Pi_{i,j} | \ydiagram[*(lightgray)]{1}\rangle = | \ydiagram[*(lightgray)]{1}\rangle\; ,
	\ee
	and it is obvious that we have 1 box to be removed. Furthermore,
	\be 
	\begin{split}
		\ytableausetup{centertableaux, smalltableaux}
		| \ydiagram[*(lightgray)]{1}\rangle = & \Pi_{1,0}  \sum_{j,k \in \mathbb{Z}} \Gamma_{j,k} | \ydiagram[*(lightgray)]{1}\rangle=
		\Pi_{1,0} \Gamma_{1,0} | \ydiagram[*(lightgray)]{1}\rangle \\
		= & \Pi_{0,1}  \sum_{j,k \in \mathbb{Z}} \Gamma_{j,k} | \ydiagram[*(lightgray)]{1}\rangle =
		\Pi_{0,1} \Gamma_{0,1} | \ydiagram[*(lightgray)]{1}\rangle \\ 
		= & \Pi_{-1,-1}  \sum_{j,k \in \mathbb{Z}} \Gamma_{j,k} | \ydiagram[*(lightgray)]{1}\rangle =
		\Pi_{-1,-1} \Gamma_{-1,-1} | \ydiagram[*(lightgray)]{1}\rangle\; .
	\end{split}
	\ee
	Therefore
	\be 
	\sum_{i,j} \Pi_{i,j}  \Gamma_{i,j}| \ydiagram[*(lightgray)]{1}\rangle = 3| \ydiagram[*(lightgray)]{1}\rangle\; ,
	\ee
	and now we have 3 places to add a box to this configuration. It is straightforward to see that this logic remains the same for 
	other configurations. 
	
	Furthermore, we also have
	\be 
	\begin{split}
	& \sum_{i,j} \Pi_{i,j}  \Gamma_{i,j}\left| \ydiagram[*(lightgray)]{1,1}\right\rangle = 3\left| \ydiagram[*(lightgray)]{1,1}\right\rangle,\\
	& \sum_{i,j} \Gamma_{i,j}  \Pi_{i,j}\left| \ydiagram[*(lightgray)]{1,1}\right\rangle = \left| \ydiagram[*(lightgray)]{1,1}\right\rangle,
	\end{split}
	\ee
	and the other 2-box states, \(| \ydiagram[*(lightgray)]{2}\rangle\) 
	and \(\left|\ytableaushort{2}*[*(lightgray)]{1}\right\rangle\), work similarly. 
	
	For 3 boxes we have two distinct cases. The first case is  
	\be 
	\begin{split}
		& \sum_{i,j} \Pi_{i,j}  \Gamma_{i,j}\left| \ydiagram[*(lightgray)]{1,1,1}\right\rangle = 3\left| \ydiagram[*(lightgray)]{1,1,1}\right\rangle\\
		& \sum_{i,j} \Gamma_{i,j}  \Pi_{i,j}\left| \ydiagram[*(lightgray)]{1,1,1}\right\rangle = \left| \ydiagram[*(lightgray)]{1,1,1}\right\rangle
	\end{split}
	\ee
	and the 3-box states \(| \ydiagram[*(lightgray)]{3}\rangle\) and \(\left|\ytableaushort{3}*[*(lightgray)]{1}\right\rangle\), 
	work similarly. 
	
	Additionally,
	\be 
	\begin{split}
		& \sum_{i,j} \Pi_{i,j}  \Gamma_{i,j}\left| \ydiagram[*(lightgray)]{1,2}\right\rangle = 4\left| \ydiagram[*(lightgray)]{1,2}\right\rangle\\
		& \sum_{i,j} \Gamma_{i,j}  \Pi_{i,j}\left| \ydiagram[*(lightgray)]{1,2}\right\rangle = 2 \left| \ydiagram[*(lightgray)]{1,2}\right\rangle,
	\end{split}
	\ee
	and the states 
	\(\left|\ytableaushort{2\none}*[*(lightgray)]{2}\right\rangle\) and 
	\(\left|\ytableaushort{\none, 2}*[*(lightgray)]{1,1}\right\rangle\) work similarly. 

	Using this logic, one may see that the Hamiltonian (\ref{eq:Hamiltonian2}) 
	is equivalent to (\ref{eq:Hamiltonian}).


	\section{Basis transformation}
	\label{app:basis}
	
	Using the characters (\ref{eq:char}) we can find the basis transformation (\ref{eq:btransf01} - \ref{eq:btransf02}) for some states. For example, the 
	1-box state is simply
	\bse  
	\be 
	\ytableausetup{centertableaux, smalltableaux}
	| \ydiagram[*(lightgray)]{1} \rangle = |(1)\rrangle\otimes |\emptyset \rangle^{\otimes} \equiv | \ydiagram[*(lightgray)]{1} \rrangle\; .
	\ee	
	The states with two boxes are
	\be 
	\ytableausetup{centertableaux, smalltableaux}
	\left|\ydiagram[*(lightgray)]{2} \right\rangle = \frac{1}{2}
	\left(\left| \ydiagram[*(lightgray)]{2} \right\rrangle +
	\left| \ydiagram[*(lightgray)]{1,1} \right\rrangle \right)
	\ee
	\be
	\nonumber
	\left| \ydiagram[*(lightgray)]{1,1} \right\rangle = -\frac{1}{2}
	\left(\left| \ydiagram[*(lightgray)]{2} \right\rrangle -
	\left| \ydiagram[*(lightgray)]{1,1} \right\rrangle \right) 
	\ee
	\be
	\nonumber
	\left|\ytableaushort[*(lightgray)]{2 }\right\rangle = |(1)\rrangle\otimes |(1)\rrangle
	\otimes |\emptyset \rangle^{\otimes} = 
	\left|\ytableaushort[*(lightgray)]{2 }\right\rrangle
	\ee
	and the states with three boxes are
	\be 
	\ytableausetup{centertableaux, smalltableaux}
	\left|\ydiagram[*(lightgray)]{3} \right\rangle = 
	\frac{1}{3}\left| \ydiagram[*(lightgray)]{3} \right\rrangle +
	\frac{1}{2}\left| \ydiagram[*(lightgray)]{1,2} \right\rrangle + 
	\frac{1}{6}\left| \ydiagram[*(lightgray)]{1,1,1} \right\rrangle
	\ee	
	\be
	\nonumber
	\left|\ydiagram[*(lightgray)]{1,1,1} \right\rangle = 
	\frac{1}{3}\left| \ydiagram[*(lightgray)]{3} \right\rrangle -
	\frac{1}{2}\left| \ydiagram[*(lightgray)]{1,2} \right\rrangle + 
	\frac{1}{6}\left| \ydiagram[*(lightgray)]{1,1,1} \right\rrangle
	\ee
	\be
	\nonumber
	\left| \ydiagram[*(lightgray)]{1,2} \right\rangle = 
	-\frac{1}{3} \left(\left| \ydiagram[*(lightgray)]{3} \right\rrangle -
	\left| \ydiagram[*(lightgray)]{1,1,1} \right\rrangle \right) \\
	\ee
	\be
	\nonumber
	|\ytableaushort{2\none}*[*(lightgray)]{2}\rangle = 
	\frac{1}{2}\left(
	|\ytableaushort{2\none}*[*(lightgray)]{2}\rrangle + \left|\ytableaushort{\none,2}*[*(lightgray)]{1,1}\right\rrangle	\right)
	\ee
	\be
	\nonumber
	\left|\ytableaushort{\none,2}*[*(lightgray)]{1,1}\right\rangle = -\frac{1}{2}\left(
	|\ytableaushort{2\none}*[*(lightgray)]{2}\rrangle - \left|\ytableaushort{\none,2}*[*(lightgray)]{1,1}\right\rrangle	
	\right)
	\ee
	\be 
	\nonumber
	|\ytableaushort{3}*[*(lightgray)]{1}\rangle= |\ytableaushort{3}*[*(lightgray)]{1}\rrangle\;.
	\ee
	\ese

	The nontrivial inverse relations are
	\bse
	\be 
	\ytableausetup{centertableaux, smalltableaux}
		\left|\ydiagram[*(lightgray)]{2} \right\rrangle = \left| \ydiagram[*(lightgray)]{2} \right\rangle - \left| \ydiagram[*(lightgray)]{1,1} \right\rangle
	\ee		
	\be
	\nonumber
		\left| \ydiagram[*(lightgray)]{1,1} \right\rrangle = \left| \ydiagram[*(lightgray)]{2} \right\rangle + \left| \ydiagram[*(lightgray)]{1,1} \right\rangle
	\ee
	and 
	\be 
	\ytableausetup{centertableaux, smalltableaux}
		\left|\ydiagram[*(lightgray)]{3} \right\rrangle = \left| \ydiagram[*(lightgray)]{3} \right\rangle - \left| \ydiagram[*(lightgray)]{1,2} \right\rangle + \left| \ydiagram[*(lightgray)]{1,1,1} \right\rangle
	\ee
	\be 
	\nonumber
		\left|\ydiagram[*(lightgray)]{1,1,1} \right\rrangle = \left| \ydiagram[*(lightgray)]{3} \right\rangle + 2 \left| \ydiagram[*(lightgray)]{1,2} \right\rangle + \left| \ydiagram[*(lightgray)]{1,1,1} \right\rangle
	\ee
	\be
	\nonumber
		\left|\ydiagram[*(lightgray)]{1,2} \right\rrangle = \left| \ydiagram[*(lightgray)]{3} \right\rangle -  \left| \ydiagram[*(lightgray)]{1,1,1} \right\rangle
	\ee
	and finally
	\be 
	\ytableausetup{centertableaux, smalltableaux}
		|\ytableaushort{2\none}*[*(lightgray)]{2}\rrangle = |\ytableaushort{2\none}*[*(lightgray)]{2}\rangle - \left|\ytableaushort{\none, 2}*[*(lightgray)]{1,1}\right\rangle
	\ee
	\be
	\nonumber
		\left|\ytableaushort{\none,2}*[*(lightgray)]{1,1}\right\rrangle = |\ytableaushort{2\none}*[*(lightgray)]{2}\rangle + \left|\ytableaushort{\none, 2}*[*(lightgray)]{1,1}\right\rangle\; .
	\ee
	\ese


	\section{Amplitudes}
	\label{app:amplitudes}

	Using formulas (\ref{eq:ampcreat}) and (\ref{eq:ampannih}) we can find the amplitude \({\cal A}_\pm\) for the first states. It is straightforward to see that
	\be 
	{\cal A}_+[\emptyset \mapsto \ydiagram[*(lightgray)]{1}] = 1, \qquad
	{\cal A}_-[\emptyset\mapsto 0] = 0\; ,
	\ee	
	where \(\chi_{\emptyset}(C[\vec{0}])=1\). The transitions \(1\mapsto 2\) boxes are
	\bse
	\be 
	\begin{split}
	{\cal A}_+& [\ydiagram[*(lightgray)]{1} \mapsto \ydiagram[*(lightgray)]{2}]= \frac{1}{Z_{(0,1,0\dots)}} 
	\chi_{(1)}[C(1,0,\dots)]\\ 
	& \times \left( 
	\chi_{(2)}[C(0,1,\dots)] + \chi_{(1,1)}[C(0,1,0\dots)]  \phantom{\frac{1}{2}} \right. \\ 
	& \left. \phantom{\frac{1}{2}} + \chi_{(1);(1)}[C((0,1,0\dots); (0,0,\dots)) ]
	\right)\\
	& = \frac{1}{2}(1-1+0)=0\; ,
	\end{split}
	\ee
	where we have used (\ref{eq:characters}) to see that 
	\be 
	\begin{split}
	\chi_{(1);(1)}& [C[(0,1,0,\dots)] \\ 
	& = \chi_{(1)}[C[(0,1,0\dots)]  \chi_{(1)}[C[(0,\dots)]=0\; .
	\end{split}
	\ee
	Similarly,
		\be 
	\begin{split}
		{\cal A}_+& \left[\ydiagram[*(lightgray)]{1} \mapsto \ydiagram[*(lightgray)]{1,1}\right]= \frac{1}{Z_{(2,0,\dots)}} 
		\chi_{(1)}[C(1,0,\dots)]\\ 
		& \times \left( 
		\chi_{(2)}[C(2,0,\dots)] + \chi_{(1,1)}[C(2,0,\dots)]  \phantom{\frac{1}{2}} \right. \\ 
		& \left. \phantom{\frac{1}{2}} + \chi_{(1);(1)}[C((2,0,\dots); (0,0,\dots)) ]
		\right) = 1\; ,
	\end{split}
	\ee
	and 
	\be 
	\begin{split}
		{\cal A}_+  [\ydiagram[*(lightgray)]{1} & \mapsto \ytableaushort{2}*[*(lightgray)]{1}] = \frac{1}{Z_{(1,0,\dots);(1,0,\dots)}} 
		\chi_{(1)}[C(1,0,\dots)]\\
		& \times \left( 
		\chi_{(2);(\emptyset)}[C(1,0,\dots);(1,0,\dots)]  \phantom{\frac{1}{2}} \right. \\ 
		&  +  \chi_{(1,1);(\emptyset)}[C(1,0,\dots);(1,0,\dots)] \\ 
		& \left. \phantom{\frac{1}{2}}  + \chi_{(1);(1)}[C((2,0,\dots); (0,0,\dots)) ] \right) = 1\; .
	\end{split}	
	\ee
	It is also immediate to see that the transition \(1\mapsto 0\)-boxes is
	\be 
	{\cal A}_-[\ydiagram[*(lightgray)]{1} \mapsto \emptyset] = 1 .
	\ee
	Therefore
	\be 
	\begin{split}
		\Gamma |\ydiagram[*(lightgray)]{1}\rrangle & = \left|\ydiagram[*(lightgray)]{1,1} \right\rrangle + ||\ytableaushort{2}*[*(lightgray)]{1}\rrangle\\
		\Pi |\ydiagram[*(lightgray)]{1}\rrangle & = |\emptyset\rrangle\; .
	\end{split}	
	\ee
	
	We can repeat the analysis above to consider the transitions \(2\mapsto 3\) boxes and \(2\mapsto 1\) box. 
	These are given by
	\be 
	\begin{split}
		{\cal A}_+[\ydiagram[*(lightgray)]{2} \mapsto \ydiagram[*(lightgray)]{3}] &= 0 \qquad 
		{\cal A}_-[\ydiagram[*(lightgray)]{2} \mapsto \ydiagram[*(lightgray)]{1}] = 0 \\
		{\cal A}_+\left[\ydiagram[*(lightgray)]{2} \mapsto \ydiagram[*(lightgray)]{1,2}\right] & = 1 \\
		{\cal A}_+[\ydiagram[*(lightgray)]{2} \mapsto \ytableaushort{2\none}*[*(lightgray)]{2}] & = 1 
	\end{split}	
	\ee
	
	\be 
	\begin{split}
		{\cal A}_+\left[\ydiagram[*(lightgray)]{1,1} \mapsto \ydiagram[*(lightgray)]{1,1,1}\right] &= 1 \qquad 
		{\cal A}_-\left[\ydiagram[*(lightgray)]{1,1} \mapsto \ydiagram[*(lightgray)]{1}\right] = 2 \\
		{\cal A}_+\left[\ydiagram[*(lightgray)]{1,1} \mapsto \ydiagram[*(lightgray)]{1,2}\right] & = 0 \\
		{\cal A}_+\left[\ydiagram[*(lightgray)]{1,1} \mapsto \ytableaushort{\none,2}*[*(lightgray)]{1,1}\right] & = 1 
	\end{split}	
	\ee
	
	\be 
	\begin{split}
		{\cal A}_+\left[\ytableaushort{2}*[*(lightgray)]{1}\mapsto \ytableaushort{2\none}*[*(lightgray)]{2}\right] & = 0 \qquad 
		{\cal A}_-\left[\ytableaushort{2}*[*(lightgray)]{1}\mapsto\ydiagram[*(lightgray)]{1}\right]= 1 \\
		{\cal A}_+\left[\ytableaushort{2}*[*(lightgray)]{1}\mapsto \ytableaushort{\none,2}*[*(lightgray)]{1,1}\right] & = 1 \\
		{\cal A}_+\left[\ytableaushort{2}*[*(lightgray)]{1}\mapsto \ytableaushort{3}*[*(lightgray)]{1 }\right] & = 1 .
	\end{split}	
	\ee
	\ese
	
	\vspace{4.0cm}
	
\clearpage
	 
	\bibliography{library.bib}

\begin{thebibliography}{28}
\expandafter\ifx\csname natexlab\endcsname\relax\def\natexlab#1{#1}\fi
\expandafter\ifx\csname bibnamefont\endcsname\relax
  \def\bibnamefont#1{#1}\fi
\expandafter\ifx\csname bibfnamefont\endcsname\relax
  \def\bibfnamefont#1{#1}\fi
\expandafter\ifx\csname citenamefont\endcsname\relax
  \def\citenamefont#1{#1}\fi
\expandafter\ifx\csname url\endcsname\relax
  \def\url#1{\texttt{#1}}\fi
\expandafter\ifx\csname urlprefix\endcsname\relax\def\urlprefix{URL }\fi
\providecommand{\bibinfo}[2]{#2}
\providecommand{\eprint}[2][]{\url{#2}}

\bibitem[{\citenamefont{von Waltershausen}(1856)}]{von1856gauss}
\bibinfo{author}{\bibfnamefont{W.}~\bibnamefont{von Waltershausen}},
  \emph{\bibinfo{title}{Gauss: zum ged{\"a}chtnis}} (\bibinfo{publisher}{S.
  Hirzel}, \bibinfo{year}{1856}),
  \urlprefix\url{https://books.google.ch/books?id=TQo\_AAAAYAAJ}.

\bibitem[{\citenamefont{Okounkov}(2003)}]{Okounkov2003}
\bibinfo{author}{\bibfnamefont{A.}~\bibnamefont{Okounkov}},
  \emph{\bibinfo{title}{The uses of random partitions}} (\bibinfo{year}{2003}),
  \eprint{math-ph/0309015}.

\bibitem[{\citenamefont{Okounkov et~al.}(2006)\citenamefont{Okounkov,
  Reshetikhin, and Vafa}}]{Okounkov:2003sp}
\bibinfo{author}{\bibfnamefont{A.}~\bibnamefont{Okounkov}},
  \bibinfo{author}{\bibfnamefont{N.}~\bibnamefont{Reshetikhin}},
  \bibnamefont{and} \bibinfo{author}{\bibfnamefont{C.}~\bibnamefont{Vafa}},
  \bibinfo{journal}{Prog. Math.} \textbf{\bibinfo{volume}{244}},
  \bibinfo{pages}{597} (\bibinfo{year}{2006}), \eprint{hep-th/0309208}.

\bibitem[{\citenamefont{Iqbal et~al.}(2008)\citenamefont{Iqbal, Nekrasov,
  Okounkov, and Vafa}}]{Iqbal:2003ds}
\bibinfo{author}{\bibfnamefont{A.}~\bibnamefont{Iqbal}},
  \bibinfo{author}{\bibfnamefont{N.}~\bibnamefont{Nekrasov}},
  \bibinfo{author}{\bibfnamefont{A.}~\bibnamefont{Okounkov}}, \bibnamefont{and}
  \bibinfo{author}{\bibfnamefont{C.}~\bibnamefont{Vafa}},
  \bibinfo{journal}{JHEP} \textbf{\bibinfo{volume}{04}}, \bibinfo{pages}{011}
  (\bibinfo{year}{2008}), \eprint{hep-th/0312022}.

\bibitem[{\citenamefont{Nekrasov and Okounkov}(2006)}]{Nekrasov:2003rj}
\bibinfo{author}{\bibfnamefont{N.}~\bibnamefont{Nekrasov}} \bibnamefont{and}
  \bibinfo{author}{\bibfnamefont{A.}~\bibnamefont{Okounkov}},
  \emph{\bibinfo{title}{{Seiberg-Witten theory and random partitions}}}
  (\bibinfo{year}{2006}), vol. \bibinfo{volume}{244}, pp.
  \bibinfo{pages}{525--596}, \eprint{hep-th/0306238}.

\bibitem[{\citenamefont{Heckman and Vafa}(2007)}]{Heckman:2006sk}
\bibinfo{author}{\bibfnamefont{J.~J.} \bibnamefont{Heckman}} \bibnamefont{and}
  \bibinfo{author}{\bibfnamefont{C.}~\bibnamefont{Vafa}},
  \bibinfo{journal}{JHEP} \textbf{\bibinfo{volume}{09}}, \bibinfo{pages}{011}
  (\bibinfo{year}{2007}), \eprint{hep-th/0610005}.

\bibitem[{\citenamefont{Jimbo and Miwa}(1983)}]{Jimbo:1983if}
\bibinfo{author}{\bibfnamefont{M.}~\bibnamefont{Jimbo}} \bibnamefont{and}
  \bibinfo{author}{\bibfnamefont{T.}~\bibnamefont{Miwa}},
  \bibinfo{journal}{Publ. Res. Inst. Math. Sci. Kyoto}
  \textbf{\bibinfo{volume}{19}}, \bibinfo{pages}{943} (\bibinfo{year}{1983}).

\bibitem[{\citenamefont{Babelon et~al.}(2003)\citenamefont{Babelon, Bernard,
  and Talon}}]{Babelon2003}
\bibinfo{author}{\bibfnamefont{O.}~\bibnamefont{Babelon}},
  \bibinfo{author}{\bibfnamefont{D.}~\bibnamefont{Bernard}}, \bibnamefont{and}
  \bibinfo{author}{\bibfnamefont{M.}~\bibnamefont{Talon}},
  \emph{\bibinfo{title}{Introduction to Classical Integrable Systems}},
  Cambridge Monographs on Mathematical Physics (\bibinfo{publisher}{Cambridge
  University Press}, \bibinfo{year}{2003}), ISBN \bibinfo{isbn}{9781139436793},
  \urlprefix\url{https://books.google.co.kr/books?id=sL0p1CoVJC0C}.

\bibitem[{\citenamefont{Dijkgraaf
  et~al.}(2009{\natexlab{a}})\citenamefont{Dijkgraaf, Orlando, and
  Reffert}}]{Dijkgraaf:2008ua}
\bibinfo{author}{\bibfnamefont{R.}~\bibnamefont{Dijkgraaf}},
  \bibinfo{author}{\bibfnamefont{D.}~\bibnamefont{Orlando}}, \bibnamefont{and}
  \bibinfo{author}{\bibfnamefont{S.}~\bibnamefont{Reffert}},
  \bibinfo{journal}{Nucl. Phys.} \textbf{\bibinfo{volume}{B811}},
  \bibinfo{pages}{463} (\bibinfo{year}{2009}{\natexlab{a}}),
  \eprint{0803.1927}.

\bibitem[{\citenamefont{Orlando et~al.}(2009)\citenamefont{Orlando, Reffert,
  and Reshetikhin}}]{Orlando:2009kd}
\bibinfo{author}{\bibfnamefont{D.}~\bibnamefont{Orlando}},
  \bibinfo{author}{\bibfnamefont{S.}~\bibnamefont{Reffert}}, \bibnamefont{and}
  \bibinfo{author}{\bibfnamefont{N.}~\bibnamefont{Reshetikhin}}
  (\bibinfo{year}{2009}), \eprint{0912.0348}.

\bibitem[{\citenamefont{Moessner and Raman}(2008)}]{Moessner2008}
\bibinfo{author}{\bibfnamefont{R.}~\bibnamefont{Moessner}} \bibnamefont{and}
  \bibinfo{author}{\bibfnamefont{K.~S.} \bibnamefont{Raman}},
  \emph{\bibinfo{title}{Quantum dimer models}} (\bibinfo{year}{2008}),
  \eprint{0809.3051}.

\bibitem[{\citenamefont{Kenyon}(2008)}]{Kenyon2008}
\bibinfo{author}{\bibfnamefont{R.}~\bibnamefont{Kenyon}},
  \bibinfo{journal}{Communications in Mathematical Physics}
  \textbf{\bibinfo{volume}{281}}, \bibinfo{pages}{675} (\bibinfo{year}{2008}),
  ISSN \bibinfo{issn}{1432-0916},
  \urlprefix\url{http://dx.doi.org/10.1007/s00220-008-0511-8}.

\bibitem[{\citenamefont{Dijkgraaf
  et~al.}(2009{\natexlab{b}})\citenamefont{Dijkgraaf, Orlando, and
  Reffert}}]{Dijkgraaf:2007yr}
\bibinfo{author}{\bibfnamefont{R.}~\bibnamefont{Dijkgraaf}},
  \bibinfo{author}{\bibfnamefont{D.}~\bibnamefont{Orlando}}, \bibnamefont{and}
  \bibinfo{author}{\bibfnamefont{S.}~\bibnamefont{Reffert}},
  \bibinfo{journal}{Adv. Theor. Math. Phys.} \textbf{\bibinfo{volume}{13}},
  \bibinfo{pages}{1255} (\bibinfo{year}{2009}{\natexlab{b}}),
  \eprint{0705.1645}.

\bibitem[{\citenamefont{Rokhsar and Kivelson}(1988)}]{Rokhsar:1988zz}
\bibinfo{author}{\bibfnamefont{D.~S.} \bibnamefont{Rokhsar}} \bibnamefont{and}
  \bibinfo{author}{\bibfnamefont{S.~A.} \bibnamefont{Kivelson}},
  \bibinfo{journal}{Phys. Rev. Lett.} \textbf{\bibinfo{volume}{61}},
  \bibinfo{pages}{2376} (\bibinfo{year}{1988}).

\bibitem[{\citenamefont{Henley}(2004)}]{Henley2003}
\bibinfo{author}{\bibfnamefont{C.}~\bibnamefont{Henley}}, \bibinfo{journal}{J.
  Phys.: Condens. Matter} \textbf{\bibinfo{volume}{16}}, \bibinfo{pages}{891}
  (\bibinfo{year}{2004}), \eprint{cond-mat/0311345}.

\bibitem[{\citenamefont{Dijkgraaf et~al.}(2010)\citenamefont{Dijkgraaf,
  Orlando, and Reffert}}]{Dijkgraaf:2009gr}
\bibinfo{author}{\bibfnamefont{R.}~\bibnamefont{Dijkgraaf}},
  \bibinfo{author}{\bibfnamefont{D.}~\bibnamefont{Orlando}}, \bibnamefont{and}
  \bibinfo{author}{\bibfnamefont{S.}~\bibnamefont{Reffert}},
  \bibinfo{journal}{Nucl. Phys. B} \textbf{\bibinfo{volume}{824}},
  \bibinfo{pages}{365} (\bibinfo{year}{2010}), \eprint{0903.0732}.

\bibitem[{\citenamefont{Ardonne et~al.}(2004)\citenamefont{Ardonne, Fendley,
  and Fradkin}}]{Ardonne:2003wa}
\bibinfo{author}{\bibfnamefont{E.}~\bibnamefont{Ardonne}},
  \bibinfo{author}{\bibfnamefont{P.}~\bibnamefont{Fendley}}, \bibnamefont{and}
  \bibinfo{author}{\bibfnamefont{E.}~\bibnamefont{Fradkin}},
  \bibinfo{journal}{Annals Phys.} \textbf{\bibinfo{volume}{310}},
  \bibinfo{pages}{493} (\bibinfo{year}{2004}), \eprint{cond-mat/0311466}.

\bibitem[{\citenamefont{Abergel et~al.}(2010)\citenamefont{Abergel, Apalkov,
  Berashevich, Ziegler, and Chakraborty}}]{Abergel_2010}
\bibinfo{author}{\bibfnamefont{D.}~\bibnamefont{Abergel}},
  \bibinfo{author}{\bibfnamefont{V.}~\bibnamefont{Apalkov}},
  \bibinfo{author}{\bibfnamefont{J.}~\bibnamefont{Berashevich}},
  \bibinfo{author}{\bibfnamefont{K.}~\bibnamefont{Ziegler}}, \bibnamefont{and}
  \bibinfo{author}{\bibfnamefont{T.}~\bibnamefont{Chakraborty}},
  \bibinfo{journal}{Advances in Physics} \textbf{\bibinfo{volume}{59}},
  \bibinfo{pages}{261} (\bibinfo{year}{2010}), ISSN \bibinfo{issn}{1460-6976},
  \urlprefix\url{http://dx.doi.org/10.1080/00018732.2010.487978}.

\bibitem[{\citenamefont{Lang}(2017)}]{lang2017}
\bibinfo{author}{\bibfnamefont{R.}~\bibnamefont{Lang}},
  \emph{\bibinfo{title}{Twists, Tilings, and Tessellations: Mathematical
  Methods for Geometric Origami}} (\bibinfo{publisher}{CRC Press},
  \bibinfo{year}{2017}), ISBN \bibinfo{isbn}{9781439873588},
  \urlprefix\url{https://books.google.nl/books?id=F91PDwAAQBAJ}.

\bibitem[{\citenamefont{Prochazka}(2016)}]{Prochazka:2015deb}
\bibinfo{author}{\bibfnamefont{T.}~\bibnamefont{Prochazka}},
  \bibinfo{journal}{JHEP} \textbf{\bibinfo{volume}{10}}, \bibinfo{pages}{077}
  (\bibinfo{year}{2016}), \eprint{1512.07178}.

\bibitem[{\citenamefont{Gaberdiel et~al.}(2017)\citenamefont{Gaberdiel,
  Gopakumar, Li, and Peng}}]{Gaberdiel:2017dbk}
\bibinfo{author}{\bibfnamefont{M.~R.} \bibnamefont{Gaberdiel}},
  \bibinfo{author}{\bibfnamefont{R.}~\bibnamefont{Gopakumar}},
  \bibinfo{author}{\bibfnamefont{W.}~\bibnamefont{Li}}, \bibnamefont{and}
  \bibinfo{author}{\bibfnamefont{C.}~\bibnamefont{Peng}},
  \bibinfo{journal}{JHEP} \textbf{\bibinfo{volume}{04}}, \bibinfo{pages}{152}
  (\bibinfo{year}{2017}), \eprint{1702.05100}.

\bibitem[{\citenamefont{Maulik and Okounkov}(2012)}]{maulik2012}
\bibinfo{author}{\bibfnamefont{D.}~\bibnamefont{Maulik}} \bibnamefont{and}
  \bibinfo{author}{\bibfnamefont{A.}~\bibnamefont{Okounkov}},
  \emph{\bibinfo{title}{Quantum groups and quantum cohomology}}
  (\bibinfo{year}{2012}), \eprint{1211.1287}.

\bibitem[{\citenamefont{Tsymbaliuk}(2017)}]{Tsymbaliuk_2017}
\bibinfo{author}{\bibfnamefont{A.}~\bibnamefont{Tsymbaliuk}},
  \bibinfo{journal}{Advances in Mathematics} \textbf{\bibinfo{volume}{304}},
  \bibinfo{pages}{583} (\bibinfo{year}{2017}), ISSN \bibinfo{issn}{0001-8708},
  \urlprefix\url{http://dx.doi.org/10.1016/j.aim.2016.08.041}.

\bibitem[{\citenamefont{Li and Yamazaki}(2020)}]{Li:2020rij}
\bibinfo{author}{\bibfnamefont{W.}~\bibnamefont{Li}} \bibnamefont{and}
  \bibinfo{author}{\bibfnamefont{M.}~\bibnamefont{Yamazaki}}
  (\bibinfo{year}{2020}), \eprint{2003.08909}.

\bibitem[{\citenamefont{Zinn-Justin}(2010)}]{Justin2008}
\bibinfo{author}{\bibfnamefont{P.}~\bibnamefont{Zinn-Justin}}, in
  \emph{\bibinfo{booktitle}{Exact Methods in Low-dimensional Statistical
  Physics and Quantum Computing}} (\bibinfo{publisher}{Oxford University
  Press}, \bibinfo{year}{2010}), \bibinfo{note}{eds Jacobsen, Ouvry, Pasquier,
  Serban, and Cugliandolo}.

\bibitem[{\citenamefont{Ketov}(1995)}]{Ketov:1995yd}
\bibinfo{author}{\bibfnamefont{S.~V.} \bibnamefont{Ketov}},
  \emph{\bibinfo{title}{Conformal field theory}} (\bibinfo{publisher}{World
  Scientific}, \bibinfo{year}{1995}).

\bibitem[{\citenamefont{Marino}(2005)}]{Marino:2005sj}
\bibinfo{author}{\bibfnamefont{M.}~\bibnamefont{Marino}},
  \bibinfo{journal}{Int. Ser. Monogr. Phys.} \textbf{\bibinfo{volume}{131}},
  \bibinfo{pages}{1} (\bibinfo{year}{2005}).

\bibitem[{\citenamefont{Fulton and Harris}(2013)}]{Fulton}
\bibinfo{author}{\bibfnamefont{W.}~\bibnamefont{Fulton}} \bibnamefont{and}
  \bibinfo{author}{\bibfnamefont{J.}~\bibnamefont{Harris}},
  \emph{\bibinfo{title}{Representation Theory: A First Course}}, Graduate Texts
  in Mathematics (\bibinfo{publisher}{Springer New York},
  \bibinfo{year}{2013}), ISBN \bibinfo{isbn}{9781461209799},
  \urlprefix\url{https://books.google.ch/books?id=6TwmBQAAQBAJ}.

\end{thebibliography}

\end{document}